%% file: kkpi0.tex
\RequirePackage{lineno}
\documentclass[a4paper,11pt]{article}
\pdfoutput=1 
\usepackage{jheppub} 
\usepackage[T1]{fontenc} 
\usepackage{graphicx}  
\usepackage{dcolumn}   
\usepackage{bm}        
\usepackage{amssymb}   
\usepackage{subfigure}
\usepackage{color}
\usepackage{xspace}
\usepackage{textpos}
\usepackage{multirow}
 \usepackage{overpic}
\include{def-com}

\def \epem {e^+e^-}

\def \kpkm {K^{+}K^{-}}

\def \kts   {K^{*+}_{2}(1430)}
\def \ktsk   {K^{*+}_{2}(1430)K^{-}}
\def \ktk   {K^{*+}_{2}(1430)K^{-}}

\def \ksk {K^{*+}(892)K^{-}}

\def \piz  {\pi^0}

\def \gev  {\mbox{GeV}}

\def \gevcc{\mbox{GeV/$c^2$}}
\def \mev  {\mbox{MeV}}

\def \mevcc{\mbox{MeV/$c^2$}}
\def \ipb  {\mbox{pb$^{-1}$}}

\leftline{Version 1 as of \today}

\title{Measurement of $\epem \to \kpkm\piz$ cross section and observation of a resonant structure}

\collaboration{BESIII Collaboration}

\abstract{
Based on $\epem$ collision data collected by the BESIII detector at the BEPCII collider at center-of-mass energies from 2.000 to 3.080~$\gev$, a partial-wave analysis is performed for the process $\epem \to \kpkm\piz$.
The Born cross section of the process $\epem \to$ $\kpkm\piz$ and its subprocesses $\epem\to\phi \piz$, $\ksk$ and $\ktk$ are measured.
The results for $\epem \to \kpkm\piz$ and $\phi \piz$ are consistent with the BaBar measurements and with improved precision. By analyzing the cross sections of the subprocesses $\epem \to$ $\ksk$ and $\ktk$, a structure with mass $M_R$ = (2208 $\pm$ 19 $\pm$ 24)~$\mevcc$ and width $\Gamma_R$ = (168 $\pm$ 24 $\pm$ 39)~$\mev$ is observed with a combined statistical significance of 7.6$\sigma$. 
The measured resonance parameters suggest it can be identified as the $\phi(2170)$, 
thus the results provide valuable input to understand the internal nature of this state.
}

\begin{document}
\maketitle
\flushbottom

\section{\label{sec:level1}Introduction}
\vspace{-0.1cm}
The vector meson state $Y(2175)$, denoted as $\phi(2170)$ by the Particle Data Group (PDG)~\cite{pdg}, 
is  one of the most interesting particles in the field of light hadron spectroscopy.
The $\phi(2170)$ was first observed by BaBar~\cite{Y2170Babar} and subsequently studied at the Belle, BESII and BESIII experiments~\cite{Y2170Babar1, Y2170belle, Y2170Bes, Y2170Bes3, 2017bes32, bes3kk, bes3phikk, bes3kkpi0pi0, bes3omegaeta, bes3phietap, bes3phieta}.
Possible interpretations of the $\phi(2170)$ state include a conventional $3^{3}S_{1}$ or $2^{3}D_{1}$ $s\bar{s}$ state~\cite{strange,2017ding,2017wang,2017afonin}, an $s\bar{s}g$ hybrid~\cite{2017ding, 2017ding2,hybrid2}, a tetraquark state~\cite{2017wang2,2017chen,2019ke,2017drenska}, a $\Lambda \bar{\Lambda}(^3S_1)$ bound state~\cite{2017zhao,2017deng,2017dong}, or a $\phi K\bar{K}$ resonance state~\cite{2017oset}.
The branching fractions of $\phi(2170)$ to various final states are critical probes to discriminate the different interpretations of $\phi(2170)$. For instance, the process $\phi(2170) \to \ktk$ has a branching fraction an order less than the branching fraction of $\phi(2170)$ $\to$ $K_{1}^{+}(1400)K^{-}$ under the hybrid hypothesis~\cite{2017ding, 2017ding2,hybrid2}, whereas under the conventional state hypothesis the two processes have branching fractions of the same order~\cite{strange,2017ding,2017wang,2017afonin}.

Experimentally,  the $\phi(2170)$ has been studied widely in $\epem$ collisions  with final states as  $\phi\eta$~\cite{babarkkpi,bes3phieta}, $\phi\eta'$~\cite{bes3phietap}, $\phi f_{0}(980)$~\cite{Y2170Babar, Y2170Babar1, Y2170belle, Y2170Bes, Y2170Bes3, 2017bes32}, $K^{+}K^{-}$~\cite{bes3kk}, $K^{*+}(892)K^{*-}(892)$~\cite{bes3kkpi0pi0}, and other $\kpkm$ states.
Among the above final states, none of them is found to be the dominant decay mode, and the products of the $\epem$ partial width and the branching fraction of each final state are all less than 10~eV.
The BaBar collaboration has studied the $\ktk$ final state~\cite{babarkkpi} by performing a Dalitz plot analysis of the process $\epem \to K_{S}^{0}K^{\pm}\pi^{\mp}$ with initial state radiation (ISR) events.   
The corresponding production cross section has been extracted and no structure has been found around 2.0~$\gev$~\cite{babarkkpi}. 
Additionally, the measured ratio of partial decay widths $\Gamma(\phi\eta)$~\cite{bes3phieta, babarkkpi} and $\Gamma(\phi\eta')$~\cite{bes3phietap} disfavors the hybrid interpretation~\cite{2017ding2,hybrid2}. However, the tension can be understood by the mechanism of hadronic transition of a strangeonium-like meson along with the $\eta-\eta'$ mixing~\cite{hybridchen}. 
Therefore, more precise measurements of $\phi(2170)$'s decay properties are desired to reveal the internal nature of  $\phi(2170)$.

In this work, we present a Partial Wave Analysis (PWA) of the process $\epem \to \kpkm\piz$ using data collected with the BESIII detector at center-of-mass (c.m.) energies ranging from 2.000 to 3.080~$\gev$ with a total integrated luminosity of 648~$\ipb$,
where the detailed values of c.m. energy  and integrated luminosities of each data set are presented in Table~\ref{data:lum}. 

\begin{table}[!htb]
\centering
\caption{Integrated luminosities of experimental data.}
 \begin{tabular}{l | c|l|c}
 \hline
 \hline

$\sqrt{s}$ (GeV)  & $\mathcal{L}$ ($\rm pb^{-1}$)  & $\sqrt{s}$ (GeV)  & $\mathcal{L}$ ($\rm pb^{-1}$)    \\
 \hline
2.000  &   10.1  &  2.396  &   66.8  \\  
2.050  &   3.4   &  2.644  &   33.6   \\ 
2.100  &   12.2  &  2.646  &   34.0    \\
2.125  &   108.5 &  2.900  &   105.0   \\
2.150  &   2.8   &  2.950  &   15.9    \\
2.175  &   10.6  &  2.981  &   16.1    \\
2.200  &   13.7  &  3.000  &   15.9    \\
2.232  &   11.8  &  3.020  &   17.3    \\
2.309  &   22.1  &  3.080  &   126.2   \\
2.386  &   22.6  &         &    \\
 
 \hline
  \hline
\end{tabular}
\label{data:lum}
\end{table}

\section{\label{sec:level2}BESIII Detector and Monte Carlo simulation}
 The BESIII detector~\cite{bes3} records symmetric $e^+e^-$ collisions
provided by the BEPCII storage ring~\cite{bepc2}, which operates with a peak luminosity of $1\times10^{33}$~cm$^{-2}$s$^{-1}$
in the center-of-mass energy range from 2.0 to 4.946~GeV.
BESIII has collected large data samples in this energy region~\cite{Ablikim:2019hff}. The cylindrical core of the BESIII detector covers 93\% of the full solid angle and consists of a helium-based multilayer drift chamber~(MDC), a plastic scintillator time-of-flight system~(TOF), and a CsI(Tl) electromagnetic calorimeter~(EMC),
which are all enclosed in a superconducting solenoidal magnet providing a 1.0~T {(0.9~T in 2012). magnetic field. The solenoid is supported by an
octagonal flux-return yoke with resistive plate counter muon identification modules interleaved with steel.
The charged-particle momentum resolution at $1~{\rm GeV}/c$ is $0.5\%$, and the $dE/dx$ resolution is $6\%$ for electrons
from Bhabha scattering. The EMC measures photon energies with a resolution of $2.5\%$ ($5\%$) at $1$~GeV in the barrel (end cap)
region. The time resolution in the TOF barrel region is 68~ps, while that in the end cap region is 110~ps.

 A Monte Carlo (MC) simulation based on {\sc Geant4}~\cite{geant4}, including the geometric description of the BESIII detector and its response, is used to optimize the event selection criteria, estimate backgrounds, and determine the detection efficiency.
 The signal MC samples are generated using the package {\sc ConExc}~\cite{conexc}, which incorporates a higher-order ISR correction.
 Background samples of the processes $\epem \to \epem$, $\mu^{+}\mu^{-}$ and $\gamma\gamma$ are generated with the {\sc Babayaga}~\cite{babayaga} generator, while $\epem \to$ hadrons and two photon events are generated by the {\sc Luarlw}~\cite{lumarlw} and {\sc Bestwogam}~\cite{bestwogam} generators, respectively.
Signal MC events are generated by using the amplitude model with parameters fixed to the PWA results.

\section{\label{sec:level3}Event selection and background analysis}
The signal process under study is $\epem \to \kpkm\piz$ with $\piz \to \gamma \gamma$. Thus, candidate events with two oppositely charged kaons and at least two photons are selected.
Charged tracks detected in the MDC are required to be within a polar angle ($\theta$) range of $|\rm{cos\theta}|<0.93$, where $\theta$ is defined with respect to the $z$-axis, and their distance of closest approach to the interaction point (IP)
must be less than 10\,cm along the $z$-axis and less than 1\,cm in the transverse plane.
Information from TOF and $dE/dx$ measurements is combined to form particle identification (PID) likelihoods for the $\pi$, $K$, and $p$ hypotheses. Each track is assigned a particle type corresponding to the hypothesis with the highest PID likelihood. 
Exactly two oppositely charged kaons are required in each event.
Photon candidates are identified using showers in the EMC. The deposited energy of each shower is more than 25~MeV in the barrel region ($|\cos \theta|< 0.80$) and more than 50~MeV in the end cap region ($0.86 <|\cos \theta|< 0.92$). 
To exclude showers induced by charged tracks, the angle between the position of each shower in the EMC and the closest extrapolated charged track is required being greater than 10$^{\circ}$.
To suppress electronic noise and showers unrelated to the event, the difference between the EMC time and the event start time is required to be within (0, 700)\,ns.

 To improve the kinematic resolution and suppress background, a four-constraint (4C) kinematic fit imposing energy-momentum conservation is carried out under the hypothesis $\epem\to\kpkm\gamma\gamma$.
 If there are more than two photons, the $\gamma\gamma$ combination with minimum $\chi^ {2}_{4C}$ is kept for further analysis.
 The candidate events are required to satisfy $\chi^{2}_{4C}$~$<$ $65$. 
 To suppress the contamination from the $\epem \to \gamma_{ISR} \phi$ process, an additional 4C kinematic fit under the hypotheses of $\epem\to\kpkm\gamma$ is performed.
 The events are discarded if the corresponding $\chi^ {2}_{4C}$  with any photon inside the event is less than the $\chi^ {2}_{4C}$ of the signal hypothesis.
Signal photons are required to have a $M_{\gamma\gamma}$ to be within the $\piz$ mass region of [0.120, 0.150] $\gevcc$.
After applying the above selection criteria, detailed studies with MC simulation indicate that the remaining background contributions are negligible.

\section{Amplitude analysis}

\subsection{Partial wave analysis method}
 Using the GPUPWA framework~\cite{PWAframe}, a PWA is performed on the surviving candidate events to identify the intermediate processes presented in $\epem \to \kpkm\piz$.
The amplitude for the $\epem \to \kpkm\piz$ decay is constructed with quasi two-body resonances using covariant tensor amplitudes~\cite{PWAtensor}.
 The intermediate states are parameterized with the relativistic Breit-Wigner (BW) functions.
To include the resolution effect for the narrow $\phi$ resonance, a Gaussian function is convolved with the BW function.
The resolution effect is negligible for the other resonances, since they have a relatively larger width.
The relative magnitudes and phases of the individual intermediate processes are determined by performing an unbinned maximum likelihood fit using MINUIT~\cite{PWA:minuit1}.
Throughout the paper, charge conjugated processes are also included by default. 

The PWA fit procedure begins by including all possible intermediate states in the PDG that match $\rm J^{PC}$ conservation in the subsequent two-body decay. These intermediate states can decay into $\kpkm$ or $K^\pm\piz$ final state.
After the fit, the statistical significance of each amplitude is evaluated by incorporating the change in likelihood and degree of freedom fits with and without the corresponding amplitude included in the fit.
Amplitudes with statistical significance $<$ 5$\sigma$ are dropped.
This procedure is repeated until a baseline solution is obtained with only amplitudes having a statistical significance $>$ 5$\sigma$.

 The above strategy is implemented individually on the data sets collected at $\sqrt{s}= 2.125$ and 2.396~$\gev$, which have the largest luminosities and yields among the nineteen data sets.
The baseline solution for data at $\sqrt{s}= 2.125$~$\gev$ includes the decay processes $\epem\to\phi\piz$, $\rho(1450)\piz$, $\phi(1680)\piz$, $\rho(1900)\piz$, $\rho_{3}(2250)\piz$, $K^{*+}(832)K^{-}$, $K^{*+}(1410)K^{-}$, $\ktsk$, $K_{3}^{*+}(1780)K^{-}$. For data at $\sqrt{s}= 2.396$~$\gev$, the intermediate process $\epem \to \rho(1450)\piz$ with a significance of 2.4$\sigma$ is excluded, while the intermediate process $\epem \to \phi\piz$ is still included in the fit to search for the possible exotic state decays to $\phi\piz$~\cite{phipiextic}. The statistical significances of all intermediate processes at the two energy points are summarized in Table~\ref{PWAtable:2125significant}.  
 The masses and widths of the $\kts$, $\phi(1680)$ and $\rho(1900)$ are determined by scanning the likelihood value in the fit.
The measured resonance parameters are consistent with the PDG values within uncertainty, except for the width of $K^{*}_{2}(1430)$ which is within 2 standard  deviations. 
The masses and widths of other intermediate states are fixed to the PDG values.
The resonant parameters and fit fractions of the intermediate states are summarized in Table~\ref{PWAtable:2125scan} and Table~\ref{PWAtable:2125fitfraction}, respectively.

\begin{table*}[htbp]
  \caption{Statistical significances of possible intermediate processes at $\sqrt{s}= 2.125$ and 2.396 GeV. }\label{PWAtable:2125significant}
  \begin{center}
  \begin{tabular}{ l | c |c}
  \hline
  \hline
   Process & Significance (2.125 GeV)                  &Significance (2.396 GeV)  \\ \hline  
   $\phi \piz$      &  18.6$\sigma$  &  2.3$\sigma$ \\
   $\rho(1450) \piz$      &  7.8$\sigma$  &  2.4$\sigma$   \\
   $\phi(1680) \piz$      &  19.5$\sigma$  &  14.9$\sigma$\\
   $\rho(1900) \piz$      &  7.2$\sigma$  &  7.4$\sigma$\\
   $\rho_{3}(2250)$       &  5.5$\sigma$  &  5.0$\sigma$\\
   \hline
   $K^{*}(892)K$          &  15.9$\sigma$  &  15.6$\sigma$ \\
   $K^{*}(1410)K$         &  5.7$\sigma$  &  5.0$\sigma$\\
   $K^{*}_{2}(1430)K$     &  35.4$\sigma$  &25.3$\sigma$\\
   $K^{*}_{3}(1780)K$     &  5.8$\sigma$  &  5.5$\sigma$\\

  \hline \hline
  \end{tabular}
  \end{center}
  \end{table*}

\begin{table}[htbp]
  \caption{Masses and widths of the intermediate states at $\sqrt{s}$ = 2.125 GeV. Due to the limited data sample size, only the statistical uncertainties are provided.}\label{PWAtable:2125scan}
  \begin{center}
  \begin{tabular}{ l | c |c | c|c} 
  \hline
  \hline
States  &   Mass ($\mevcc$)     &   Width (MeV)    &  PDG Mass ($\mevcc$)  &  PDG Width (MeV) \\
\hline
   $K^{*}_{2}(1430)$      & 1428 $\pm$ 2   & 107 $\pm$ 4    & 1427.3 $\pm$1.5  &  100.0  $\pm$  2.1   \\
   $\phi(1680)$           & 1673 $\pm$ 5   & 172 $\pm$ 8   & 1680 $\pm$ 20    &  150 $\pm$  50   \\
   $\rho(1900)$            & 1880 $\pm$ 10  & 69  $\pm$ 15   & 1860 $-$  1910    &  10 $-$ 160      \\
   $\phi$            & fixed           &   fixed       &   1019.5 $\pm$ 0.02 & 4.2 $\pm$ 0.01 \\
   $\rho(1450)$           & fixed           &   fixed       & 1465 $\pm$ 25       & 400 $\pm$ 60 \\
   $\rho_{3}(2250)$       & fixed           &   fixed       & 2248$^{+17+59}_{-17-5}$ & 185$^{+31+17}_{-26-103}$\\
   $K^{*}(892)$           & fixed           &   fixed       & 891.7 $\pm$ 0.3      & 50.8 $\pm$ 0.9 \\
   $K^{*}(1410)$           & fixed           &   fixed       & 1414 $\pm$ 15       & 232 $\pm$ 21 \\
   $K^{*}_{3}(1780)$           & fixed           &   fixed       & 1776 $\pm$ 7    & 159 $\pm$ 21 \\ 
\hline\hline
\end{tabular}
\end{center}
\end{table}

\begin{table*}[htbp]
  \caption{Fit fractions of possible intermediate processes at $\sqrt{s}= 2.125$ and 2.396 GeV. }\label{PWAtable:2125fitfraction}
  \begin{center}
  \begin{tabular}{ l | c |c}
  \hline
  \hline
   Process & Fraction (\%) (2.125 GeV)  &Fraction (\%) (2.396 GeV)  \\ \hline  
   $\phi \piz$            &  1.8  $\pm$ 0.4  &  0.6  $\pm$ 0.3 \\
   $\rho(1450) \piz$      &  3.8  $\pm$ 0.7  &  --\\
   $\phi(1680) \piz$      &  14.6 $\pm$ 2.3  &  15.5 $\pm$ 2.1 \\
   $\rho(1900) \piz$      &  2.1  $\pm$ 0.3  &  2.7  $\pm$ 1.0 \\
   $\rho_{3}(2250)$       &  0.9  $\pm$ 0.5  &  0.9  $\pm$ 0.6 \\
   \hline
   $K^{*}(892)K$          &  2.8  $\pm$  0.3  & 9.3  $\pm$ 1.1 \\
   $K^{*}(1410)K$         &  1.1  $\pm$  0.8  & 4.2  $\pm$ 1.3 \\
   $K^{*}_{2}(1430)K$     &  73.0 $\pm$  3.7  & 66.6 $\pm$ 2.8 \\
   $K^{*}_{3}(1780)K$     &  1.3  $\pm$  0.5  & 2.7  $\pm$ 1.4 \\

  \hline \hline
  \end{tabular}
  \end{center}
  \end{table*}

The invariant mass spectra and angular distributions in data and fit results are shown in Figs.~\ref{PWAfiguure2125:1} and~\ref{PWAfiguure2396:1}, respectively.
The $\chi^{2}/nbin$ value is displayed on each figure to demonstrate the goodness of fit, where $nbin$ is the number of bins of each figure and $\chi^{2}$ is defined as:
 \begin{equation}
      \chi^{2} = \sum_{i=1}^{nbin}\frac{(n_i-\nu_i)^{2}}{n_i},
 \label{eq0}
 \end{equation}
where $n_i$ and $\nu_i$ are the number of events for the data and the fit projections in the ith bin of each figure, respectively.

 \begin{figure*}[!htp]
 \centering
 \begin{overpic}[width=0.32\textwidth, height=0.27\textwidth]{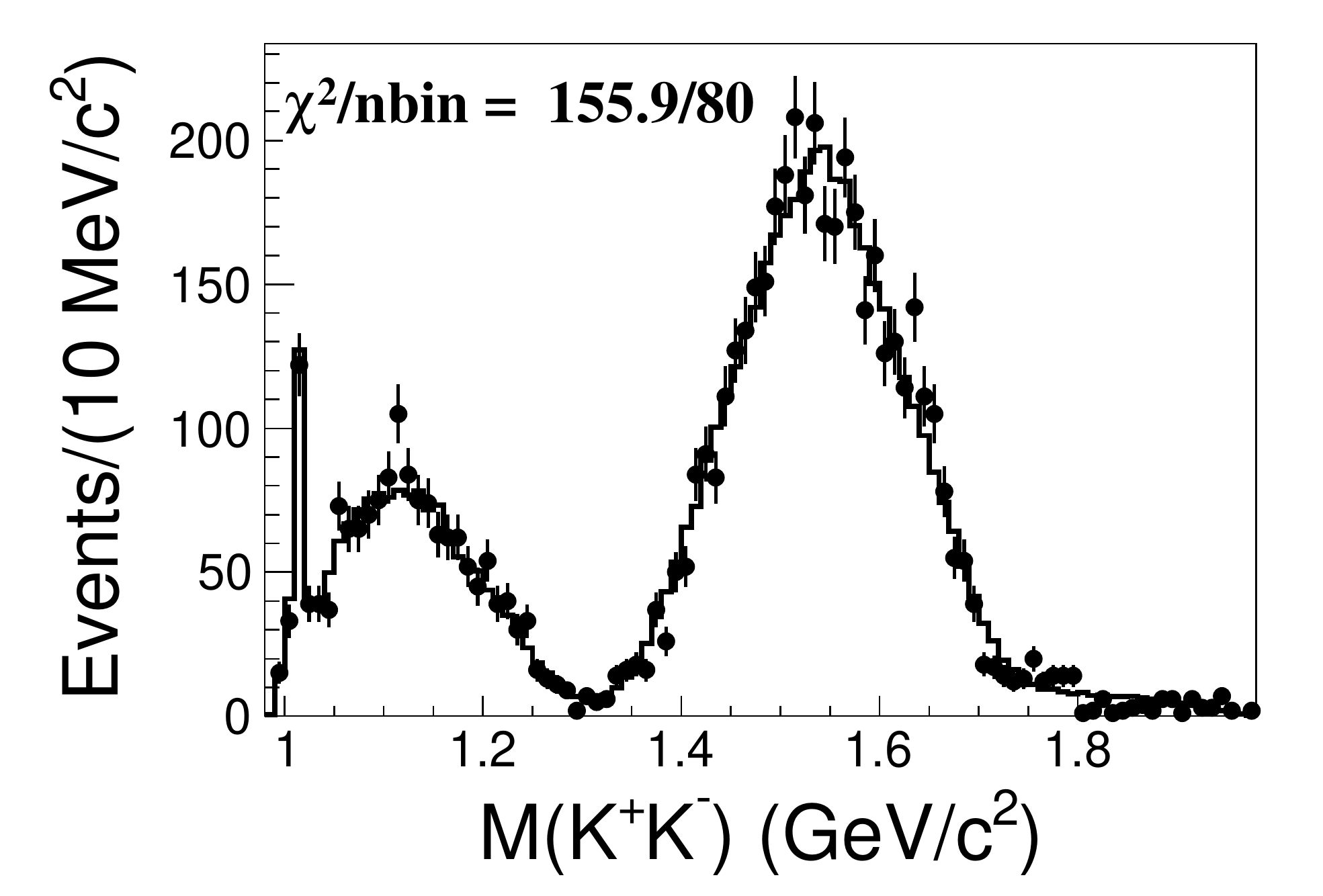}
 \put(65,70){ (a)}
 \end{overpic}
 \begin{overpic}[width=0.32\textwidth, height=0.27\textwidth]{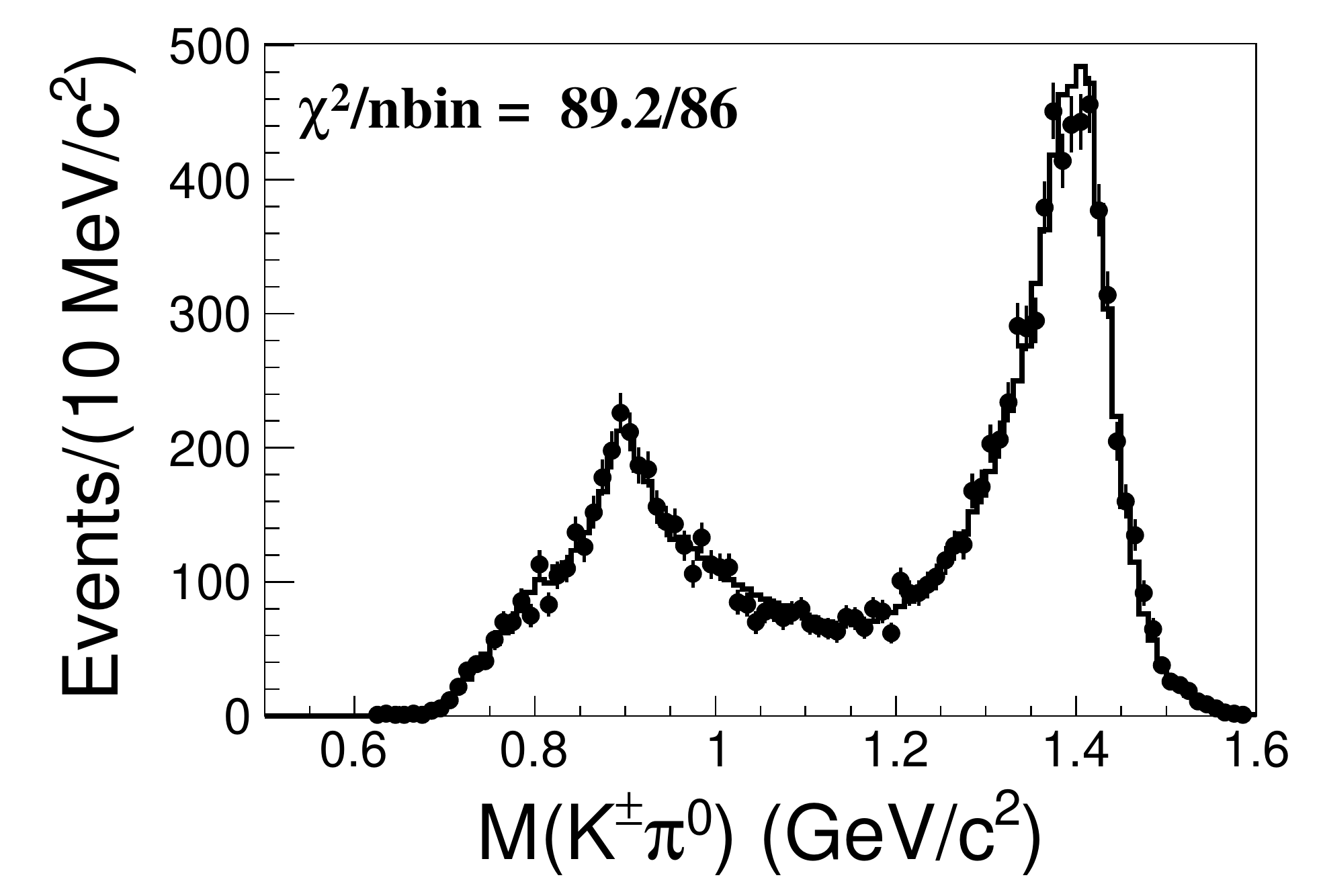}
 \put(60,70){ (b)}
 \end{overpic}

 \begin{overpic}[width=0.32\textwidth, height=0.27\textwidth]{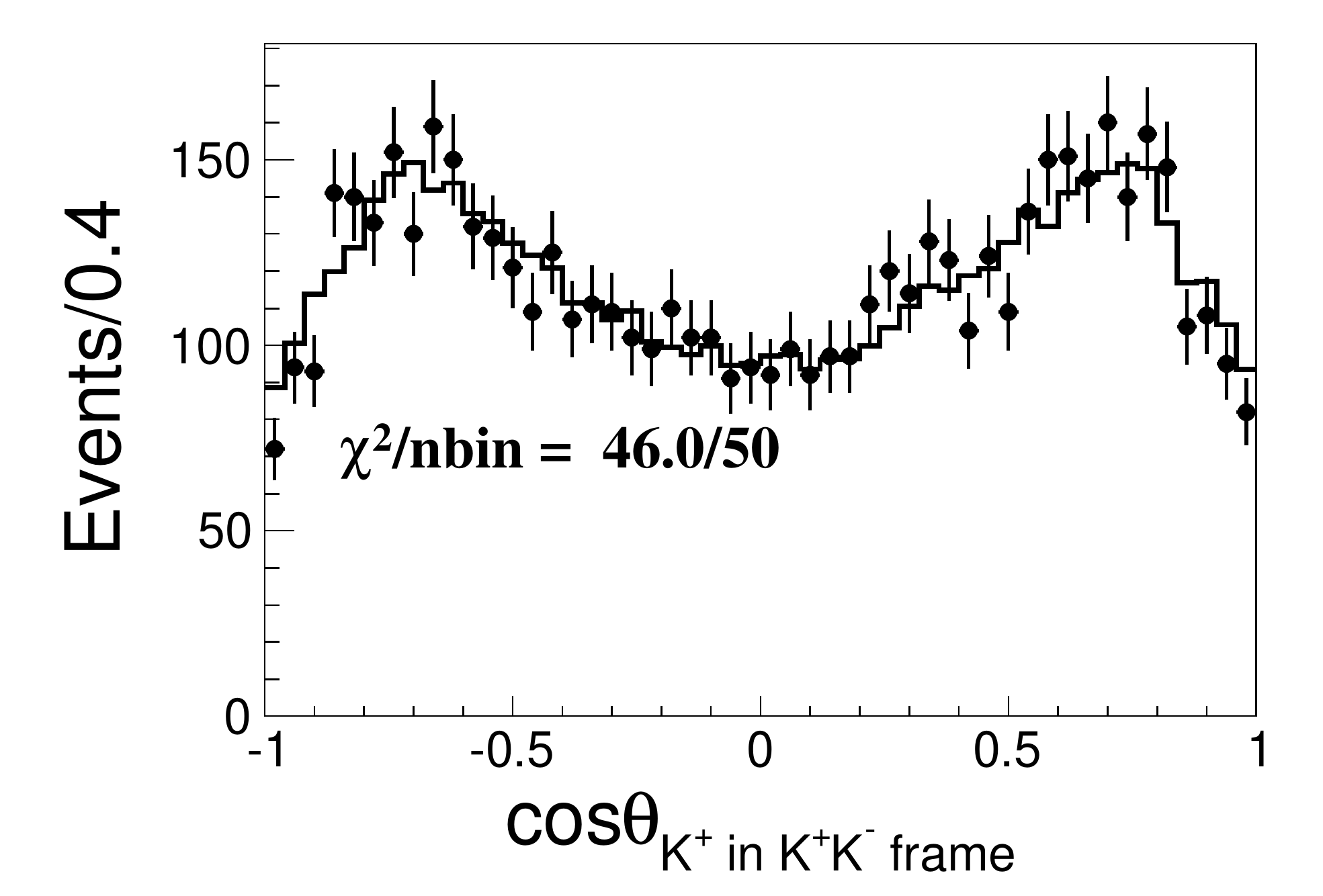}
 \put(60,70){ (c)}
 \end{overpic}
 \begin{overpic}[width=0.32\textwidth, height=0.27\textwidth]{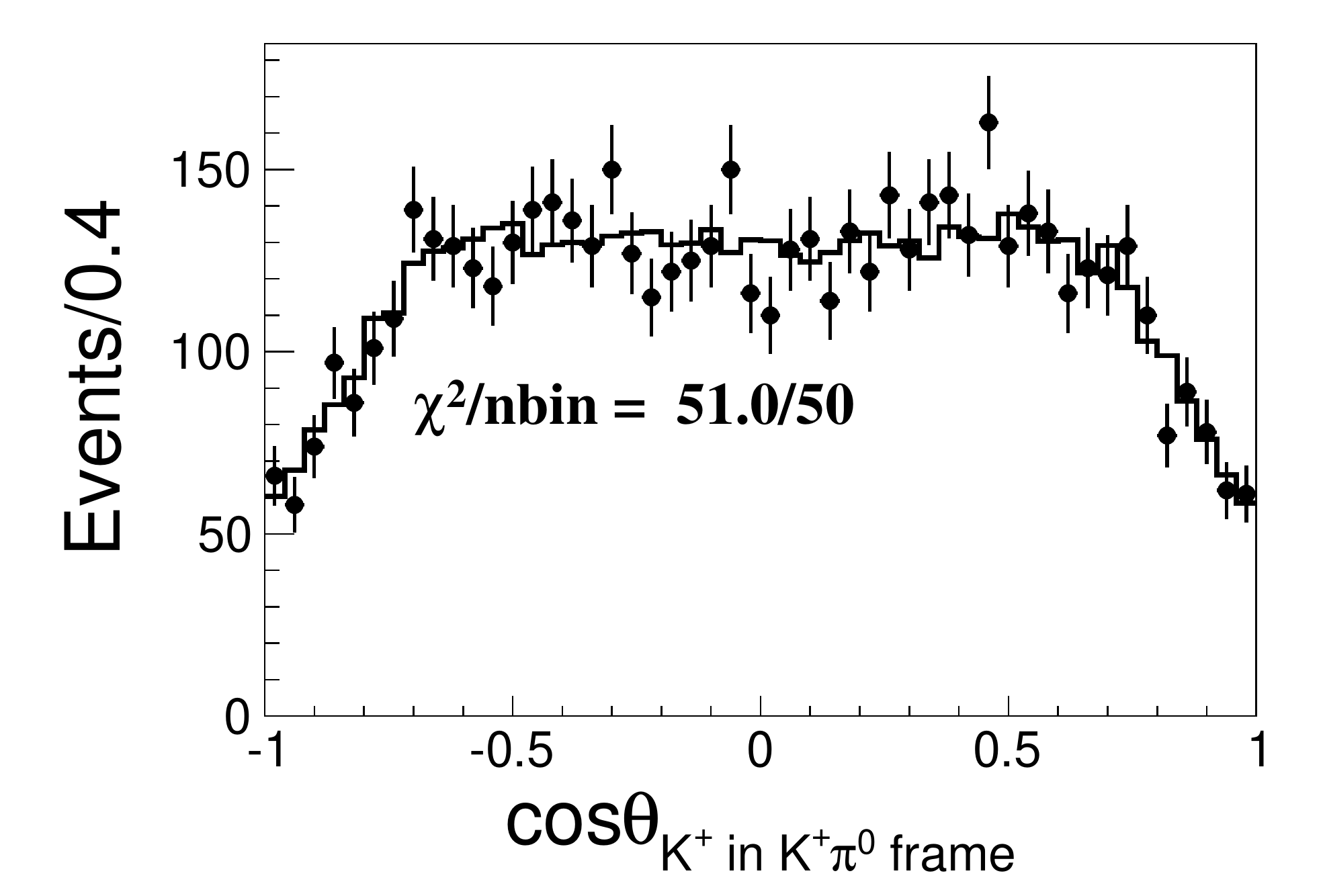}
 \put(75,70){ (d)}
 \end{overpic}
 \begin{overpic}[width=0.32\textwidth, height=0.27\textwidth]{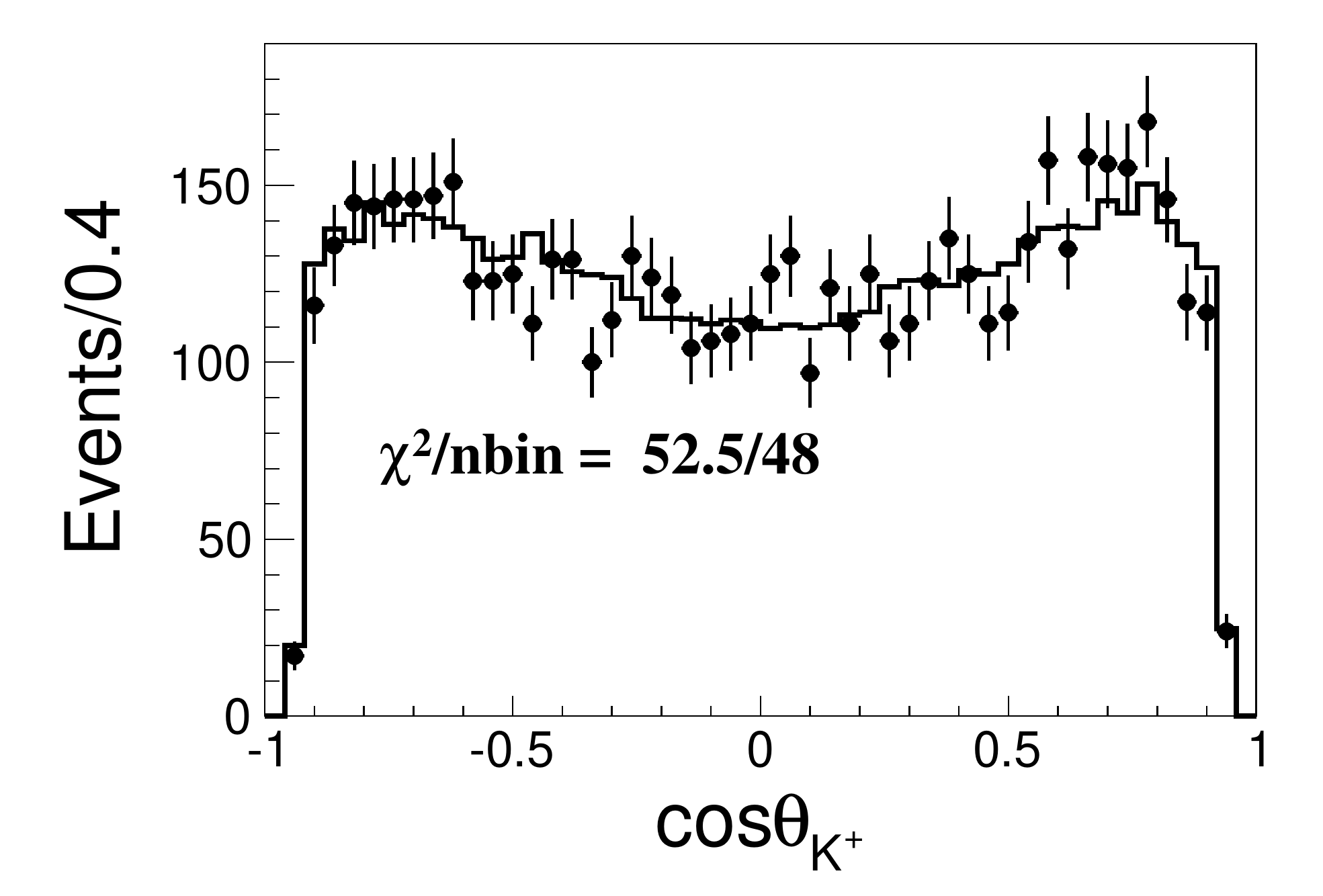}
 \put(60,70){ (e)}
 \end{overpic}
 \caption{ (a) Invariant mass distribution of $\kpkm$; (b) invariant mass distribution of $K^{\pm}\piz$; (c) cos$\theta$ distribution of $\it K^{+}$ in the $\kpkm$ rest frame; (d) cos$\theta$ distribution of $\it K^{+}$ in the $K^{+}\pi^{0}$ rest frame; (e) cos$\theta$ distribution of $K^{+}$ in the c.m. frame at $\sqrt{s} = $ 2.125 GeV. $\theta$ is polar angle with respect to the $z$-axis. Dots with error bars are data, and the curves are the fit results.}
 \label{PWAfiguure2125:1}
\end{figure*}

  \begin{figure*}[!htp]
 \centering
 \begin{overpic}[width=0.32\textwidth, height=0.27\textwidth]{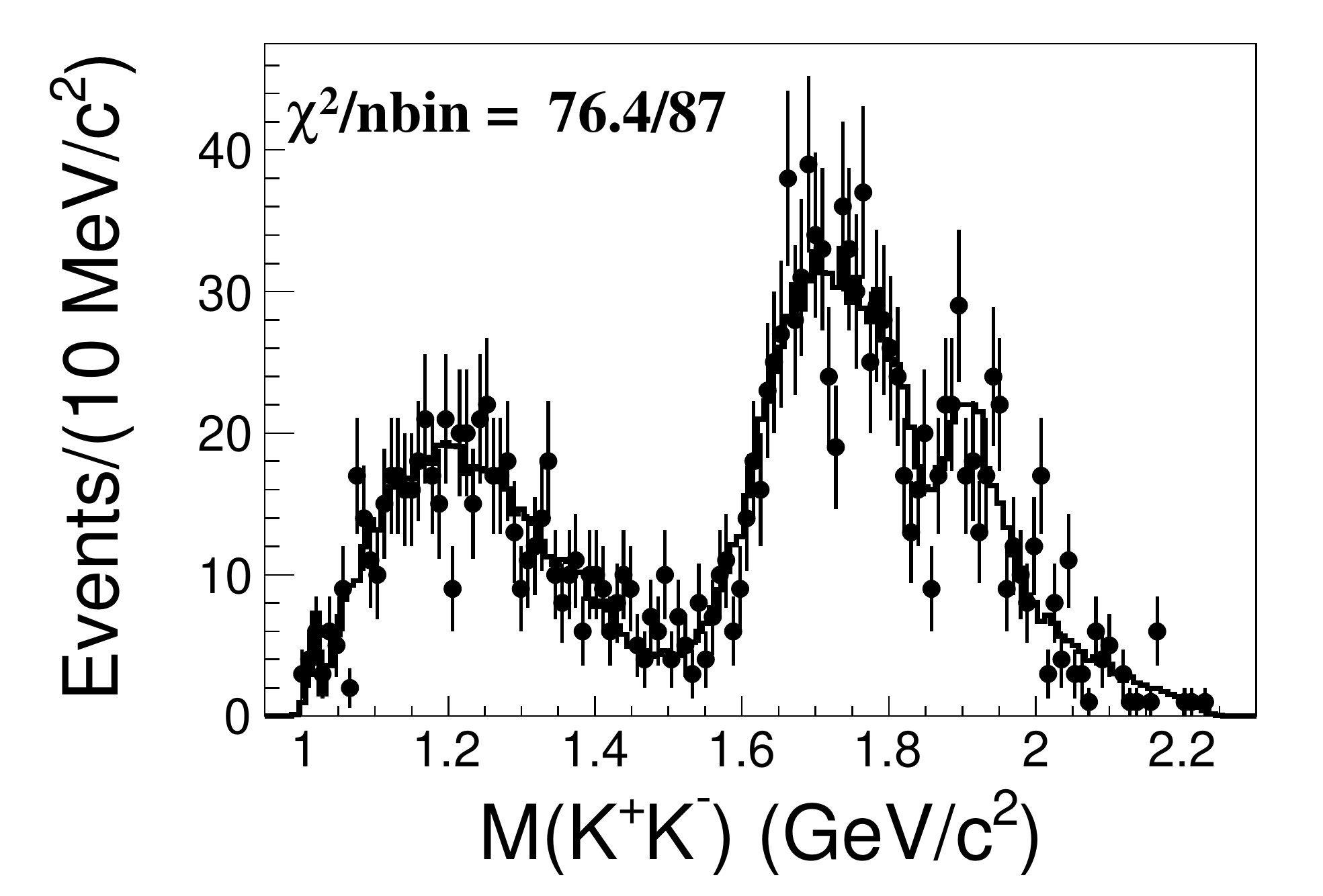}
 \put(70,70){ (a)}
 \end{overpic}
 \begin{overpic}[width=0.32\textwidth, height=0.27\textwidth]{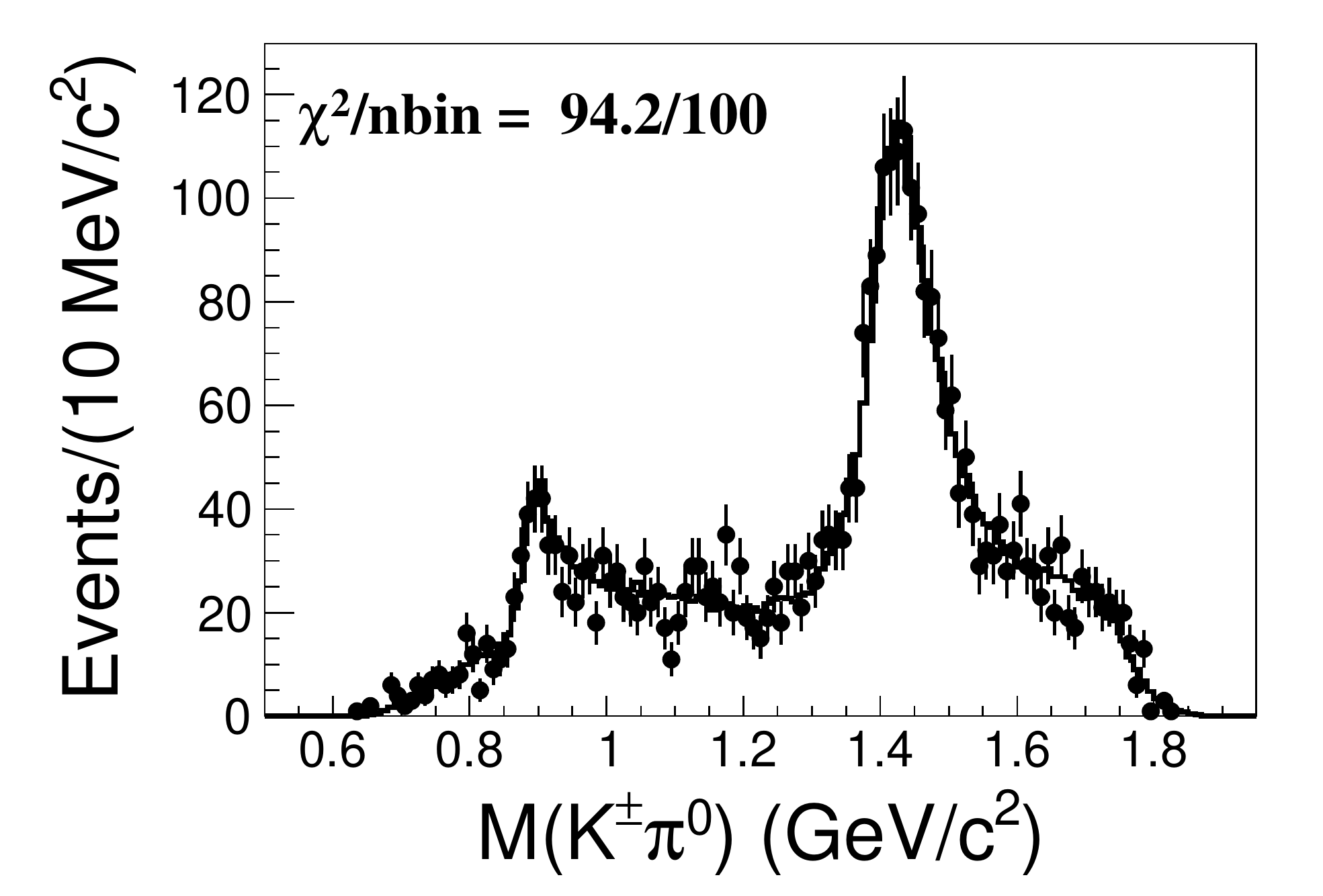}
 \put(70,70){ (b)}
 \end{overpic}

 \begin{overpic}[width=0.32\textwidth, height=0.27\textwidth]{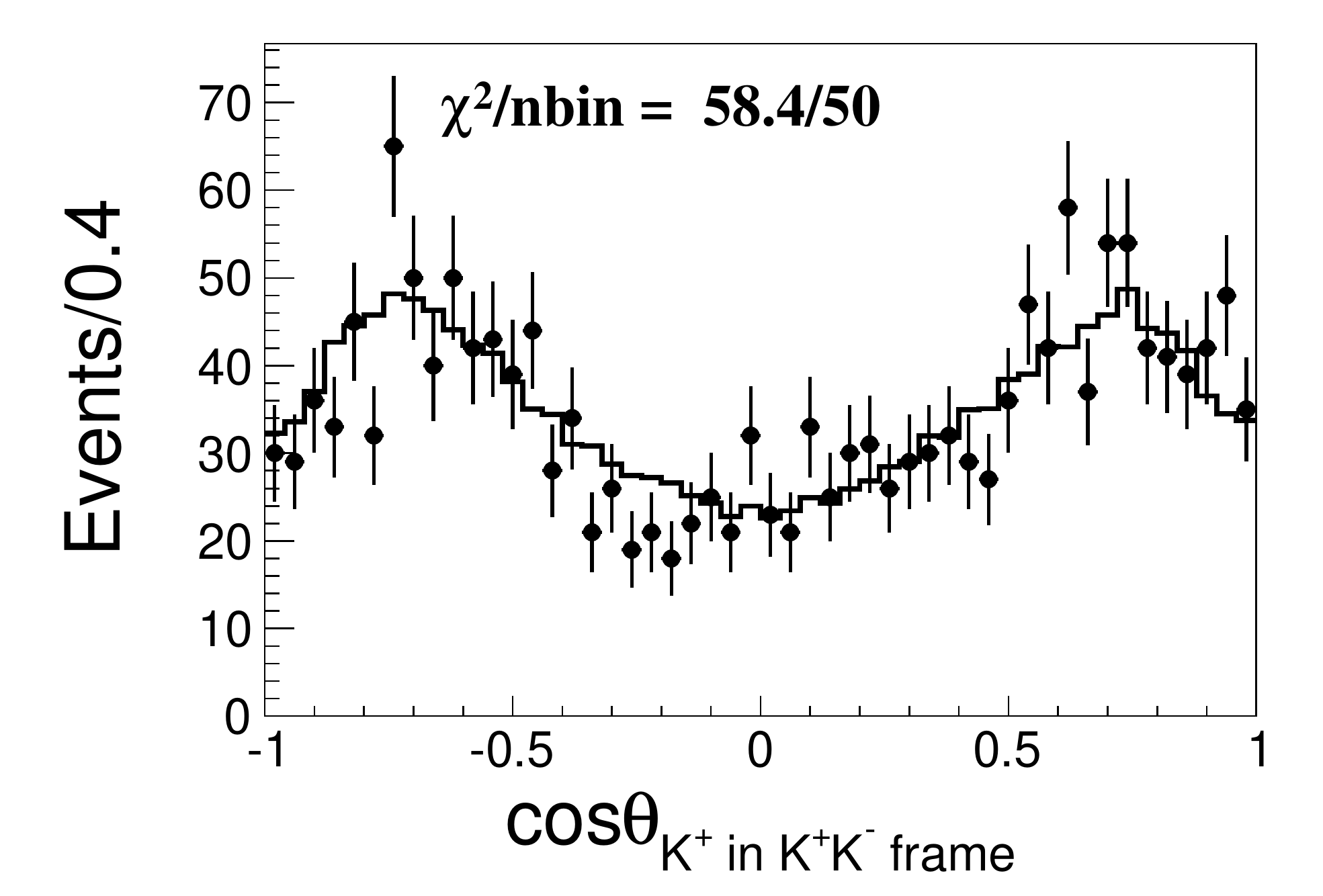}
 \put(70,70){ (c)}
 \end{overpic}
 \begin{overpic}[width=0.32\textwidth, height=0.27\textwidth]{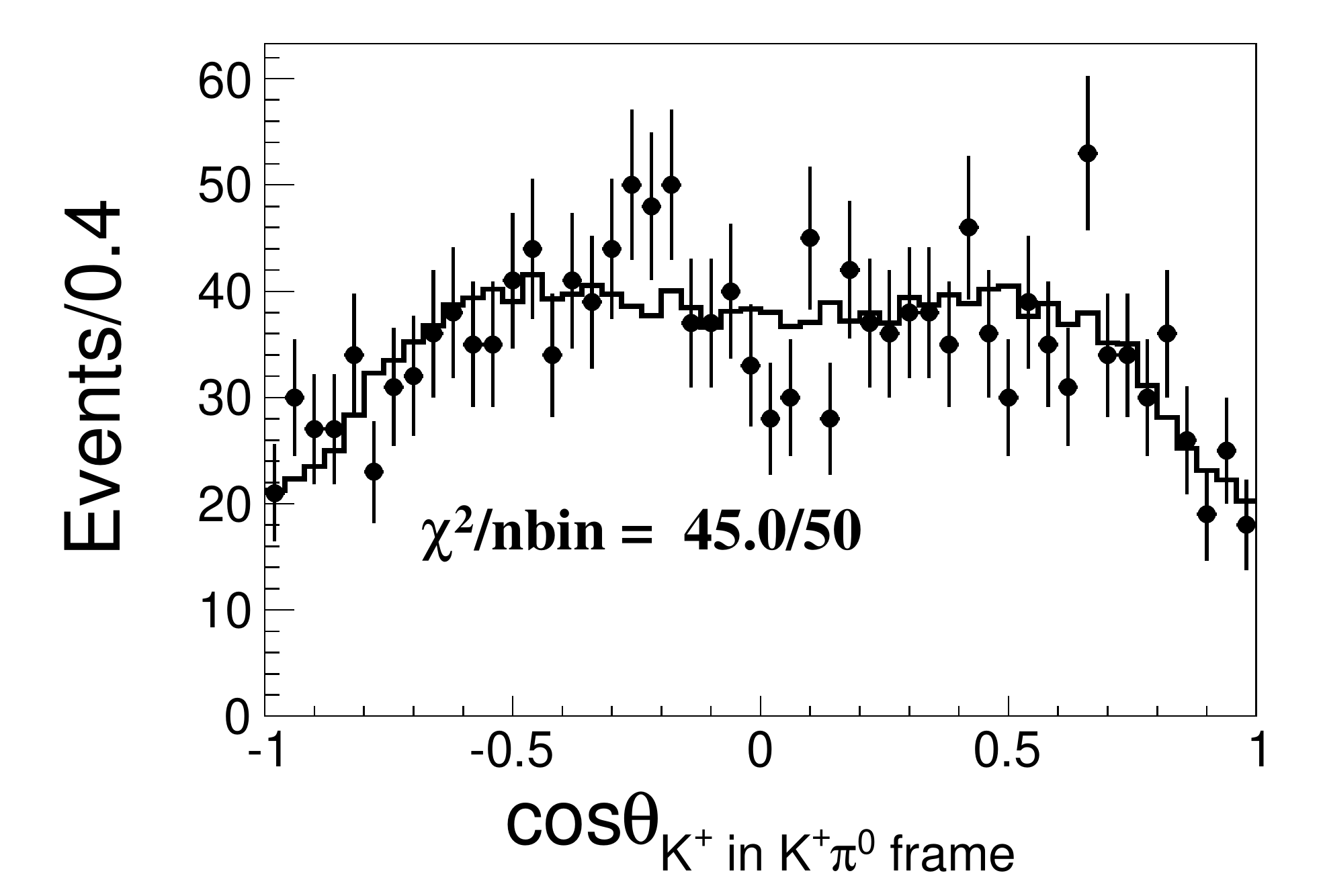}
 \put(70,70){ (d)}
 \end{overpic}
 \begin{overpic}[width=0.32\textwidth, height=0.27\textwidth]{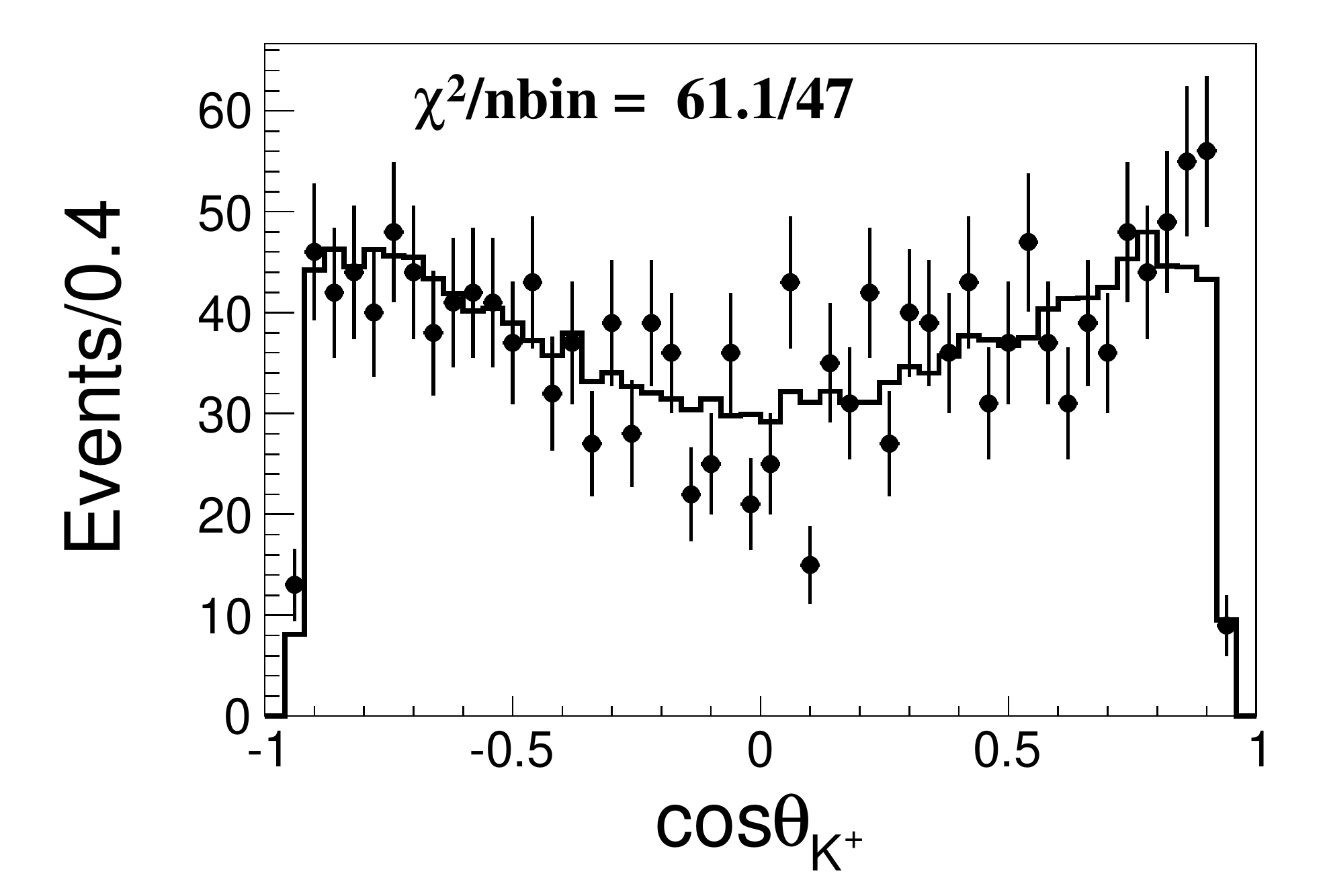}
 \put(70,70){ (e)}
 \end{overpic}
 \caption{ (a) Invariant mass distribution of $\kpkm$; (b) invariant mass distribution of $K^{\pm}\piz$; (c) cos$\theta$ distribution of $\it K^{+}$ in the $\kpkm$ rest frame; (d) cos$\theta$ distribution of $\it K^{+}$ in the $K^{+}\pi^{0}$ rest frame; (e) cos$\theta$ distribution of $K^{+}$ in the c.m. frame at $\sqrt{s} = $ 2.396 GeV. $\theta$ is polar angle with respect to the $z$-axis. Dots with error bars are data, and the curves are the fit results.}
 \label{PWAfiguure2396:1}
\end{figure*}

The other seventeen data samples have limited statistics, so they are fitted using the resonances from the baseline solutions obtained with $\sqrt{s}=2.125~\gev$ (for $\sqrt{s}=$ 2.000, 2.100, 2.175, 2.200 and 2.232 GeV, hereafter called group I) and $\sqrt{s}=$2.396 GeV (for the other data samples, hereafter named group II). For the two groups, the parameters of intermediate states are fixed, and the magnitude and phase of each process are float.

\subsection{Cross section measurement}

 The total Born cross section for $\epem\to \kpkm\piz$ is obtained at the individual c.m.~energy by using:
 \begin{equation}
     \sigma^{\rm B}=\frac{N^{\rm sig}}{\mathcal{L}_{\rm int}\cdot\frac{1}{|1-\Pi|^{2}}\cdot(1+\delta)^{\rm r}\cdot\mathcal{B}r\cdot\epsilon},
     \label{equation}
 \end{equation}
where $N^{\rm sig}$ is the corresponding signal yield, which is the number of surviving events due to the negligible background;
 $\mathcal{L}_{\rm int}$ is the integrated luminosity;
 $(1+\delta)^{\rm r}$ is the ISR correction factor obtained from QED calculations~\cite{VR,conexc} by incorporating the input cross section from this analysis iteratively;
 $\frac{1}{|1-\Pi|^{2}}$ is the vacuum polarization factor taken from QED calculations~\cite{VP};
 $\epsilon$ is the detection efficiency obtained from weighting MC simulation according to the PWA results;
$\mathcal{B}r$ is the branching ratio of the decay $\piz \to \gamma\gamma$ quoted from the PDG~\cite{pdg}.
Meanwhile, the Born cross sections for the intermediate processes  are obtained with the same approach, individually, while the signal yield $N^{\rm sig}$  is replaced with the product of
the total number of surviving events and the corresponding fraction relative to the total signal yields obtained according to the PWA results, and $\mathcal{B}r$ is replaced with the product of the branching ratio of the decay $\piz \to \gamma\gamma$ and that of the intermediate states quoted from the PDG~\cite{pdg}.
The number of events of the intermediate processes, $\epem \to \phi\piz$, $\ksk$ and $\ktsk$, are extracted in the PWA fit to calculate the cross section. 
The measured cross sections as well as the signal yields are summarized in Tables~\ref{cro:kkpi0}-\ref{cro:ksk}, separately for the process $\epem \to \kpkm\piz$ and for each individual intermediate process.

\begin{table}[!htb]
\centering
\caption{ The c.m. energy, detection efficiency, radiative correction factor, vacuum polarization factor, measured cross section for the process $\epem \to K^{+}K^{-} \piz$, where the first uncertainties are statistical, and the second are systematic.}
    \begin{tabular}{l | c|c|c|c|c}
      \hline
  \hline

$\sqrt{s}$ (GeV) & N$^{\rm sig}$      & $\epsilon$ (\%)   & $(1+\delta)^{\rm r}$ & $\frac{1}{|1-\Pi|^{2}}$   &$\sigma^{\rm B}$ (pb)  \\
 \hline
2.000  &    400.0 $\pm$ 20.0  &      23.4  &   0.979  &   1.037  &  168.4 $\pm$ 8.4 $\pm$ 6.6 \\
2.050  &   208.0 $\pm$ 14.4  &       24.5  &   0.962  &   1.038  &   256.3 $\pm$ 17.8 $\pm$ 9.7 \\
2.100  &    712.0 $\pm$ 26.7  &      25.2  &   0.975  &   1.039  &   232.5 $\pm$ 8.7 $\pm$ 9.1 \\
2.125  &     5894.0 $\pm$ 76.8  &    25.2  &   0.988  &   1.039  &   212.2 $\pm$ 2.8 $\pm$ 8.1 \\
2.150  &   152.0 $\pm$ 12.3  &       25.4  &   1.000  &   1.040  &   204.5 $\pm$ 16.6 $\pm$ 8.0 \\
2.175  &    504.0 $\pm$ 22.4  &      25.6  &   1.014  &   1.040  &   178.7 $\pm$ 8.0 $\pm$ 7.0 \\
2.200  &    632.0 $\pm$ 25.1  &      24.9  &   1.029  &   1.040  &   175.1 $\pm$ 7.0 $\pm$ 6.8 \\
2.232  &    520.0 $\pm$ 22.8  &      24.6  &   1.048  &   1.041  &   165.5 $\pm$ 7.3 $\pm$ 6.3 \\
2.309  &    800.0 $\pm$ 28.3  &      23.7  &   1.091  &   1.041  &   136.2 $\pm$ 4.8 $\pm$ 5.3 \\
2.386  &    636.0 $\pm$ 25.2  &      22.8  &   1.128  &   1.041  &   106.5 $\pm$ 4.2 $\pm$ 4.2 \\
2.396  &    1736.0 $\pm$ 41.7  &     22.9  &   1.133  &   1.041  &   97.4 $\pm$ 2.3 $\pm$ 3.8 \\
2.644  &    512.0 $\pm$ 22.6  &      21.5  &   1.205  &   1.039  &   57.2 $\pm$ 2.5 $\pm$ 2.2 \\
2.646  &    560.0 $\pm$ 23.7  &      21.5  &   1.203  &   1.039  &   62.1 $\pm$ 2.6 $\pm$ 2.4 \\
2.900  &     864.0 $\pm$ 29.4  &     21.4  &   1.218  &   1.033  &   31.0 $\pm$ 1.1 $\pm$ 1.2 \\
2.950  &    112.0 $\pm$ 10.6  &      21.5  &   1.225  &   1.029  &   26.3 $\pm$ 2.5 $\pm$ 1.1 \\
2.981  &    120.0 $\pm$ 11.0  &      21.1  &   1.230  &   1.025  &   28.4 $\pm$ 2.6 $\pm$ 1.1 \\
3.000  &    124.0 $\pm$ 11.1  &      20.6  &   1.236  &   1.021  &   30.5 $\pm$ 2.7 $\pm$ 1.2 \\
3.020  &    108.0 $\pm$ 10.4  &      21.0  &   1.242  &   1.014  &   23.9 $\pm$ 2.3 $\pm$ 1.0 \\
3.080  &     824.0 $\pm$ 28.7  &     19.6  &   1.309  &   0.915  &   28.2 $\pm$ 1.0 $\pm$ 1.1 \\

 \hline
  \hline
\end{tabular}
\label{cro:kkpi0}
\end{table}

\begin{table}[!htb]
\centering
\caption{ The c.m. energy, detection efficiency, radiative correction factor, vacuum polarization factor, measured cross section for the process $\epem \to \phi\pi^{0}$, where the first uncertainties are statistical, and the second are systematic.}
    \begin{tabular}{l | c|c|c|c|c}
      \hline
  \hline
$\sqrt{s}$ (GeV)  & N$^{\rm sig}$      & $\epsilon$ (\%)  & $(1+\delta)^{\rm r}$ & $\frac{1}{|1-\Pi|^{2}}$    &$\sigma^{\rm B}$ (pb)  \\
 \hline
2.000  &     32.6 $\pm$ 10.3  &     17.6  &   1.604  &   1.037  &   22.6 $\pm$ 7.1 $\pm$ 2.3 \\
2.050  &    11.3 $\pm$ 4.7  &      16.6  &   1.734  &   1.038  &   23.3 $\pm$ 9.7 $\pm$ 2.3 \\
2.100  &     20.9 $\pm$ 11.8  &      17.5  &   1.671  &   1.039  &   11.7 $\pm$ 6.6 $\pm$ 1.2 \\
2.125  &      106.4 $\pm$ 22.0  &     18.1  &   1.621  &   1.039  &   6.6 $\pm$ 1.4 $\pm$ 0.7 \\
2.150  &    9.0 $\pm$ 3.9  &     18.8  &   1.572  &   1.040  &   21.3 $\pm$ 9.1 $\pm$ 2.1 \\
2.175  &     16.9 $\pm$ 5.5  &     19.6  &   1.546  &   1.040  &   10.4 $\pm$ 3.4 $\pm$ 1.0 \\
2.200  &     11.0 $\pm$ 5.3  &     19.3  &   1.592  &   1.040  &   5.2 $\pm$ 2.5 $\pm$ 0.5 \\
2.232  &     6.8 $\pm$ 3.2  &      15.9  &   2.027  &   1.041  &   3.5 $\pm$ 1.6 $\pm$ 0.3 \\
2.309  &     0.2 $\pm$ 0.2  &     9.8  &     3.335  &   1.041  &   0.1 $\pm$ 0.1 $\pm$ 0.1 \\
2.386  &     10.5 $\pm$ 3.8  &      10.0  &   3.104  &   1.041  &   3.0 $\pm$ 1.1 $\pm$ 0.3 \\
2.396  &     10.8 $\pm$ 5.5  &      11.0  &   2.860  &   1.041  &   1.0 $\pm$ 0.5 $\pm$ 0.1 \\

 \hline
  \hline
\end{tabular}
\label{cro:phipi0}
\end{table}

\begin{table}[!htb]
\centering
\caption{ The c.m. energy, detection efficiency, radiative correction factor, vacuum polarization factor, measured cross section for the process $\epem \to \ktsk$, where the first uncertainties are statistical, and the second are systematic.}
    \begin{tabular}{l | c|c|c|c|c}
      \hline
  \hline

$\sqrt{s}$ (GeV)   & N$^{\rm sig}$      & $\epsilon$ (\%)  & $(1+\delta)^{\rm r}$ & $\frac{1}{|1-\Pi|^{2}}$     &$\sigma^{\rm B}$ (pb)  \\
 \hline
2.000  &    138.5 $\pm$ 38.8  &     29.7  &   0.769  &   1.037  &   347.6 $\pm$ 97.5 $\pm$ 35.3 \\
2.050  &   57.1 $\pm$ 20.2  &      30.5  &   0.797  &   1.038  &   406.8 $\pm$ 144.1  $\pm$ 41.0\\
2.100  &    413.2 $\pm$ 98.6  &      30.5  &   0.846  &   1.039  &   764.1 $\pm$ 182.3 $\pm$ 77.1\\
2.125  &     4304.1 $\pm$ 227.1  &     30.0  &   0.877  &   1.039  &   873.3 $\pm$ 46.1 $\pm$ 89.0 \\
2.150  &   132.3 $\pm$ 25.5  &      29.4  &   0.902  &   1.040  &   1015.9 $\pm$ 196.0 $\pm$ 103.5\\
2.175  &    417.8 $\pm$ 53.1  &      29.2  &   0.925  &   1.040  &   845.7 $\pm$ 107.6 $\pm$ 86.2\\
2.200  &    441.4 $\pm$ 48.3  &     28.6  &   0.944  &   1.040  &   691.7 $\pm$ 75.6 $\pm$ 70.5\\
2.232  &    377.4 $\pm$ 42.6  &      28.3  &   0.963  &   1.041  &   676.2 $\pm$ 76.3 $\pm$ 70.0\\
2.309  &    548.8 $\pm$ 44.6  &      27.1  &   0.998  &   1.041  &   531.9 $\pm$ 43.3 $\pm$ 29.2\\
2.386  &    364.1 $\pm$ 30.8  &      26.3  &   1.023  &   1.041  &   347.5 $\pm$ 29.4 $\pm$ 19.1\\
2.396  &    1156.4 $\pm$ 56.1  &      26.2  &   1.025  &   1.041  &   373.0 $\pm$ 18.1 $\pm$ 20.5\\
2.644  &    339.4 $\pm$ 27.3  &      23.6  &   1.102  &   1.039  &   225.0 $\pm$ 18.1 $\pm$ 12.4\\
2.646  &    413.5 $\pm$ 27.3  &      23.4  &   1.103  &   1.039  &   272.5 $\pm$ 18.0 $\pm$ 15.0\\
2.900  &     516.7 $\pm$ 29.5  &      20.4  &   1.233  &   1.033  &   114.0 $\pm$ 6.5 $\pm$ 6.3\\
2.950  &    45.0 $\pm$ 10.0  &     19.6  &   1.268  &   1.029  &   66.6 $\pm$ 14.8  $\pm$ 3.6\\
2.981  &    52.8 $\pm$ 10.8  &     19.4  &   1.286  &   1.025  &   77.7 $\pm$ 15.9 $\pm$ 4.2\\
3.000  &    71.3 $\pm$ 11.1  &     19.0  &   1.299  &   1.021  &   107.4 $\pm$ 16.7 $\pm$ 5.9\\
3.020  &    36.4 $\pm$ 13.2  &     18.9  &   1.309  &   1.014  &   50.6 $\pm$ 18.4 $\pm$ 2.8\\
3.080  &     379.4 $\pm$ 23.7  &      17.6  &   1.352  &   0.915  &   83.0 $\pm$ 5.2 $\pm$ 4.6\\

 \hline
  \hline
\end{tabular}
\label{cro:k2sk}
\end{table}

\begin{table}[!htp]
\centering
\caption{ The c.m. energy, detection efficiency, radiative correction factor, vacuum polarization factor, measured cross section for the process $\epem \to K^{*+}(892)K^{-}$, where the first uncertainties are statistical, and the second are systematic.}
    \begin{tabular}{l | c|c|c|c|c}
      \hline
  \hline

$\sqrt{s}$ (GeV)   & N$^{\rm sig}$      & $\epsilon$ (\%)   & $(1+\delta)^{\rm r}$ & $\frac{1}{|1-\Pi|^{2}}$     &$\sigma^{\rm B}$ (pb)  \\
 \hline
2.000  &    23.1 $\pm$ 10.3  &    22.5  &   0.990  &   1.037  &   30.0 $\pm$ 13.4 $\pm$ 3.3 \\
2.050  &   22.7 $\pm$ 9.3  &      23.6  &   0.972  &   1.038  &   86.6 $\pm$ 35.6 $\pm$ 9.5 \\
2.100  &    22.3 $\pm$ 17.1  &    23.8  &   0.985  &   1.039  &   22.9 $\pm$ 17.6 $\pm$ 2.5 \\
2.125  &     163.2 $\pm$ 19.8  &  23.7  &   0.997  &   1.039  &   18.6 $\pm$ 2.3 $\pm$ 2.0 \\
2.150  &   5.0 $\pm$ 5.3  &       23.5  &   1.009  &   1.040  &   21.5 $\pm$ 23.0 $\pm$ 2.4 \\
2.175  &    13.1 $\pm$ 7.0  &     23.6  &   1.024  &   1.040  &   14.9 $\pm$ 7.9 $\pm$ 1.6 \\
2.200  &    56.0 $\pm$ 13.0  &    23.3  &   1.038  &   1.040  &   49.4 $\pm$ 11.4 $\pm$ 5.4 \\
2.232  &    36.5 $\pm$ 10.8  &    23.2  &   1.057  &   1.041  &   36.8 $\pm$ 10.8 $\pm$ 4.0 \\
2.309  &    76.0 $\pm$ 13.2  &    22.5  &   1.099  &   1.041  &   40.6 $\pm$ 7.1 $\pm$ 1.7 \\
2.386  &    69.5 $\pm$ 14.2  &    22.1  &   1.134  &   1.041  &   35.9 $\pm$ 7.4 $\pm$ 1.5 \\
2.396  &    161.6 $\pm$ 20.1  &   22.1  &   1.139  &   1.041  &   28.1 $\pm$ 3.5 $\pm$ 1.2 \\
2.644  &    29.6 $\pm$ 9.1  &     21.2  &   1.211  &   1.039  &   10.0 $\pm$ 3.1 $\pm$ 0.4 \\
2.646  &    24.5 $\pm$ 12.4  &    21.2  &   1.210  &   1.039  &   8.2 $\pm$ 4.2 $\pm$ 0.3 \\
2.900  &     55.2 $\pm$ 10.5  &   21.2  &   1.221  &   1.033  &   6.0 $\pm$ 1.1 $\pm$ 0.2 \\
2.950  &    8.0 $\pm$ 4.4  &      20.9  &   1.227  &   1.029  &   5.8 $\pm$ 3.2 $\pm$ 0.2 \\
2.981  &    2.3 $\pm$ 4.1  &      20.9  &   1.233  &   1.025  &   1.7 $\pm$ 2.9 $\pm$ 0.1 \\
3.000  &    1.2 $\pm$ 0.9  &      20.8  &   1.236  &   1.021  &   0.9 $\pm$ 0.6 $\pm$ 0.1 \\
3.020  &    1.8 $\pm$ 3.5  &      20.7  &   1.240  &   1.014  &   1.2 $\pm$ 2.4 $\pm$ 0.1 \\
3.080  &     173.1 $\pm$ 15.6  &  19.6  &   1.271  &   0.915  &   18.3 $\pm$ 1.7 $\pm$ 0.8 \\

 \hline
  \hline
\end{tabular}
\label{cro:ksk}
\end{table}

\subsection{Systematic uncertainties for the intermediate states}
 Two categories of systematic uncertainties are considered in the measurement of the Born cross sections.

 The first category includes those associated with the luminosity, track detection, PID, kinematic fit, ISR correction, and the branching fractions of intermediate states.
 The uncertainty associated with the integrated luminosity is 1\% at each energy point~\cite{lum}.
 The uncertainty of the detection efficiency is 1\% for each charged track~\cite{trackerror} and photon~\cite{photonerror}, individually.
 The PID efficiency uncertainty is 1.0\% for each charged track~\cite{trackerror}.
 The uncertainty related to the kinematic fit is estimated by correcting the helix parameters of the simulated charged tracks to match the resolution~\cite{helixsys}.
The uncertainty associated with the ISR and VP effect is obtained with the accuracy of the radiation function, which is about 0.5\%~\cite{VP}, and has a contribution from the cross section lineshape, which is estimated by varying the model parameters of the fit to the cross section. All parameters are randomly varied within their uncertainties and the resulting parametrization of the lineshape is used to recalculate $(1+\delta)^{r}\epsilon$ and the corresponding cross section. This procedure is repeated five hundred times and the standard deviation of the resulting cross section is considered as systematic uncertainty.
The uncertainty of the $\piz$ invariant-mass requirement is evaluated by tuning the MC sample for the $\piz$ mass resolution according to data at $\sqrt{s} = $ 2.125 GeV.
 The systematic uncertainties from the branching ratios of intermediate states in the subsequent decays are taken from the PDG~\cite{pdg} and propagated.

 The second category of uncertainties are associated with the PWA fit.
Fits with alternative scenarios  are performed, and the changes of signal yields are taken as systematic uncertainties.
 Uncertainties associated with  the BW parametrization are estimated by replacing the constant-width BW with the mass-dependent width.
Uncertainties associated with the resonance parameters, which are taken from the PDG and fixed in the fit, are estimated by performing alternative fits with the added constraints that each resonance parameter follows a Gaussian distribution with a width equal to its uncertainty. 
 One thousand fits are performed, and the resulting relative deviations of the signal yields are taken as systematic uncertainties.
 Uncertainties associated with the additional resonances are estimated by alternative fits including the components $K^{*}(1680)K$ or the $\rho(1700)\piz$, which resulted being the most significant, even if with a significance less than 5$\sigma$ obtained from data.
 Uncertainties due to the barrier factor are estimated by varying the radius of the centrifugal barrier from 0.7 to 1.0 fm and considering the difference in $\sigma^{\rm B}$ as the uncertainty.
Uncertainties associated with the MC mode for $\epem \to \kpkm\piz$ cross section are estimated by the alternative PWA mode including all the components with a significance more than 3$\sigma$. 

 In the above procedure, the uncertainties associated with the barrier factor, resonance parametrization and additional resonances are strongly affected by the statistics. Thus, those uncertainties of data with $\sqrt{s}$=2.125~$\gev$ are assigned to the group I data, while those of data with $\sqrt{s}$=2.396~$\gev$ are assigned to the group II data.
For the process $\epem \to \phi\pi^{0}$, due to the limited statistics at $\sqrt{s}$=2.396~$\gev$, the uncertainties obtained at $\sqrt{s}$=2.125~$\gev$ are assigned to all the data sets. 

 Assuming all the sources of systematic uncertainties as independent, the total uncertainties are the quadratic sums of the individual values, as shown in Tables~\ref{Sys:kkpi0}-\ref{Sys:ksk}, where the sources of the uncertainties tagged with `*' are assumed to be 100\% correlated among c.m. energies.

\begin{table}[!htp]
\centering
\scriptsize
\caption{ Systematic uncertainties (in \%) of $e^{+}e^{-} \to \kpkm\piz$ at each energy point, where the sources of the uncertainties tagged with ``*'' are assumed to be 100\% correlated among each energy point.}
    \begin{tabular}{l | cccccccccc}
      \hline
  \hline

Data set  & $\mathcal{L}$*  &  Pho.*  &     Track*  &      PID*  & $\pi^{0}$ Mass* & Kim.   &    ISR  &     Br* & PWA mode  &  Sum \\
2.000      &  1.0    &    2.0     &     2.0      &      2.0     &     0.3    &       0.73    &     1.0      &           0.03  & 0.9   &  3.9\\
2.050     &  1.0    &    2.0     &     2.0      &      2.0     &     0.3    &       0.71    &     0.8      &           0.03   & 0.9   &  3.8\\
2.100      &  1.0    &    2.0     &     2.0      &      2.0     &     0.3    &       0.75    &     1.1      &           0.03  &0.9    &  3.9\\
2.125    &  1.0    &    2.0     &     2.0      &      2.0     &     0.3    &       0.08    &     0.8      &           0.03    &0.9    &  3.8\\
2.150    &  1.0    &    2.0     &     2.0      &      2.0     &     0.3    &       0.64    &     1.0      &           0.03    &0.9    &  3.8\\
2.175    &  1.0    &    2.0     &     2.0      &      2.0     &     0.3    &       0.69    &     0.9      &           0.03    &0.9    &  3.9\\
2.200    &  1.0    &    2.0     &     2.0      &      2.0     &     0.3    &       0.71    &     0.9      &           0.03    &0.9    &  3.9\\
2.232   &  1.0    &    2.0     &     2.0      &      2.0     &     0.3    &       0.68    &     0.8      &           0.03    &0.9     &  3.9\\
2.309   &  1.0    &    2.0     &     2.0      &      2.0     &     0.3    &       0.66    &     0.8      &           0.03    &1.2     &  3.9\\
2.386   &  1.0    &    2.0     &     2.0      &      2.0     &     0.3    &       0.61    &     0.8      &           0.03    & 1.2    &  3.9\\
2.396    &  1.0    &    2.0     &     2.0      &      2.0     &     0.3    &       0.26    &     0.6      &           0.03   & 1.2    &  3.9\\
2.644    &  1.0    &    2.0     &     2.0      &      2.0     &     0.3    &       0.61    &     0.7      &           0.03   & 1.2  &  3.9\\
2.646    &  1.0    &    2.0     &     2.0      &      2.0     &     0.3    &       0.60    &     0.7      &           0.03   & 1.2  &  3.9\\
2.900    &  1.0    &    2.0     &     2.0      &      2.0     &     0.3    &       0.52    &     1.0      &           0.03    &1.2 &  4.0\\
2.950    &  1.0    &    2.0     &     2.0      &      2.0     &     0.3    &       0.53    &     1.0      &           0.03    &1.2 &  4.0\\
2.981    &  1.0    &    2.0     &     2.0      &      2.0     &     0.3    &       0.52    &     1.0      &           0.03    &1.2 &  4.0\\
3.000    &  1.0    &    2.0     &     2.0      &      2.0     &     0.3    &       0.55    &     1.1      &           0.03    &1.2 &  4.0\\
3.020    &  1.0    &    2.0     &     2.0      &      2.0     &     0.3    &       0.53    &     1.0      &           0.03    &1.2 &  4.0\\
3.080    &  1.0    &    2.0     &     2.0      &      2.0     &     0.3    &       0.53    &     1.0      &           0.03    &1.2 &  4.0\\
 \hline
  \hline
\end{tabular}
\label{Sys:kkpi0}
\end{table}

\begin{table}[hbp]
\centering
\scriptsize
\caption{ Systematic uncertainties (in \%) of $e^{+}e^{-} \to \phi \pi^{0}$ at each energy point, where the sources of the uncertainties tagged with ``*" are assumed to be 100\% correlated among each energy point.}
    \begin{tabular}{l | cccccccccccc}
     \hline
  \hline
Data set  & $\mathcal{L}$*  &  Pho.*  &     Track*  &      PID*  & $\pi^{0}$ Mass* & Kim.   &    ISR  &   Res.para &      Barrer   &        Add.Res     &        Br*   &  Sum \\

2.000     &  1.0  & 2.0    &    2.0   &   2.0  &      0.3    &     1.04   &   0.89   &  1.1   &     5.6     &        9.3     &     1.0  & 11.6\\
2.050    &  1.0  & 2.0    &    2.0   &   2.0  &      0.3    &     1.03   &   0.87   &  1.1   &     5.6     &        9.3     &     1.0  & 11.6\\
2.100     &  1.0  & 2.0    &    2.0   &   2.0  &      0.3    &     0.93   &   0.90   &  1.1   &     5.6     &        9.3     &     1.0  & 11.6\\
2.125   &  1.0  & 2.0    &    2.0   &   2.0  &      0.3    &     0.98   &   0.50   &  1.1   &     5.6     &        9.3     &     1.0  & 11.6\\
2.150   &  1.0  & 2.0    &    2.0   &   2.0  &      0.3    &     0.99   &   0.81   &  1.1   &     5.6     &        9.3     &     1.0  & 11.6\\
2.175   &  1.0  & 2.0    &    2.0   &   2.0  &      0.3    &     0.96   &   0.85   &  1.1   &     5.6     &        9.3     &     1.0  & 11.6\\
2.200   &  1.0  & 2.0    &    2.0   &   2.0  &      0.3    &     0.93   &   0.87   &  1.1   &     5.6     &        9.3     &     1.0  & 11.6\\
2.232  &  1.0  & 2.0    &    2.0   &   2.0  &      0.3    &     0.89   &   0.84   &  1.1   &     5.6     &        9.3     &     1.0  & 11.6\\
2.309  &  1.0  & 2.0    &    2.0   &   2.0  &      0.3    &     0.80   &   0.84   &  1.1   &     5.6     &        9.3     &     1.0  & 11.6\\
2.386  &  1.0  & 2.0    &    2.0   &   2.0  &      0.3    &     0.83   &   0.79   &  1.1   &     5.6     &        9.3     &     1.0  & 11.6\\
2.396   &  1.0  & 2.0    &    2.0   &   2.0  &      0.3    &     0.85   &   0.56   &  1.1   &     5.6     &        9.3     &     1.0  & 11.6\\
 \hline
  \hline
\end{tabular}
\label{Sys:phipi0}
\end{table}

\begin{table}[htbp]
\centering
\scriptsize
\caption{ Systematic uncertainties (in \%) of $e^{+}e^{-} \to K_{2}^{*+}(1430)K^{-}$ at each energy point, where the sources of the uncertainties tagged with ``*'' are assumed to be 100\% correlated among each energy point.}
    \begin{tabular}{l | cccccccccccc}
      \hline
  \hline
Data set  & $\mathcal{L}$*  &  Pho.*  &     Track*  &      PID*  & $\pi^{0}$ Mass* & Kim.   &    ISR  &   Res.para &      Barrer   &        Add.Res     &        Br*   &  Sum \\
2.000   &   1.0  &   2.0   &     2.0    &      2.0    &    0.3   &     0.85  &     0.6  &   3.4     &        5.5     &             6.3   &         2.4 &   10.2\\
2.050   &  1.0  &   2.0   &     2.0    &      2.0    &    0.3   &     0.96  &     0.6  &   3.4     &        5.5     &             6.3   &         2.4 &   10.2\\
2.100    &  1.0  &   2.0   &     2.0    &      2.0    &    0.3   &     0.76  &     0.7  &   3.4     &        5.5     &            6.3    &        2.4  &  10.2\\
2.125  &  1.0  &   2.0   &     2.0    &      2.0    &    0.3   &     1.32  &     0.6  &   3.4     &        5.5     &             6.3   &         2.4 &   10.3\\
2.150  &  1.0  &   2.0   &     2.0    &      2.0    &    0.3   &     0.80  &     0.6  &   3.4     &        5.5     &             6.3   &         2.4 &   10.2\\
2.175  &  1.0  &   2.0   &     2.0    &      2.0    &    0.3   &     0.83  &     0.5  &   3.4     &        5.5     &             6.3   &         2.4 &   10.2\\
2.200  &  1.0  &   2.0   &     2.0    &      2.0    &    0.3   &     0.76  &     0.5  &   3.4     &        5.5     &             6.3   &         2.4 &   10.2\\
2.232 &  1.0  &   2.0   &     2.0    &      2.0    &    0.3   &     0.78  &     0.5  &   3.4     &        5.5     &             6.3   &         2.4 &   10.3\\
2.309 &  1.0  &   2.0   &     2.0    &      2.0    &    0.3   &     0.73  &     0.5  &   1.5     &        1.0     &             1.2   &         2.4 &   5.4\\
2.386 &  1.0  &   2.0   &     2.0    &      2.0    &    0.3   &     0.75  &     0.5  &   1.5     &        1.0     &             1.2   &         2.4 &   5.4\\
2.396  &  1.0  &   2.0   &     2.0    &      2.0    &    0.3   &     0.71  &     0.5  &   1.5     &        1.0     &             1.2   &         2.4 &   5.4\\
2.644  &  1.0  &   2.0   &     2.0    &      2.0    &    0.3   &     0.69  &     0.5  &   1.5     &        1.0     &             1.2   &         2.4 &   5.5\\
2.646  &  1.0  &   2.0   &     2.0    &      2.0    &    0.3   &     0.63  &     0.5  &   1.5     &        1.0     &             1.2   &         2.4 &   5.4\\
2.900  &  1.0  &   2.0   &     2.0    &      2.0    &    0.3   &     0.58  &     0.5  &   1.5     &        1.0     &             1.2   &         2.4 &   5.5 \\
2.950  &  1.0  &   2.0   &     2.0    &      2.0    &    0.3   &     0.63  &     0.5  &   1.5     &        1.0     &             1.2   &         2.4 &   5.5\\
2.981  &  1.0  &   2.0   &     2.0    &      2.0    &    0.3   &     0.56  &     0.5  &   1.5     &        1.0     &             1.2   &         2.4 &   5.6\\
3.000  &  1.0  &   2.0   &     2.0    &      2.0    &    0.3   &     0.58  &     0.5  &   1.5     &       1.0      &            1.2    &        2.4  &  5.7\\
3.020  &  1.0  &   2.0   &     2.0    &      2.0    &    0.3   &     0.56  &     0.5  &   1.5     &       1.0      &            1.2    &        2.4  &  5.7\\
3.080  &  1.0  &   2.0   &     2.0    &      2.0    &    0.3   &     0.60  &     0.6  &   1.5     &        1.0     &             1.2   &         2.4 &   5.7 \\

 \hline
  \hline
\end{tabular}
\label{Sys:k2sk}
\end{table}

\begin{table}[htbp]
\centering
\scriptsize
\caption{ Systematic uncertainties (in \%) of $e^{+}e^{-} \to K^{*+}(892)K^{-}$ at each energy point, where the sources of the uncertainties tagged with ``*'' are assumed to be 100\% correlated among each energy point.}
    \begin{tabular}{l | cccccccccccc}
      \hline
  \hline

Data set  & L*  &  Pho.*  &     Track*  &      PID*  & $\pi^{0}$ Mass* & Kim.   &    ISR  &   Res.para &      Barrer   &        Add.Res     &        Br*   &  Sum \\

2.000    &   1.0    &   2.0    &    2.0     &     2.0   &     0.3   &      0.93   &   1.0  &   0.8   &    4.4   &   9.3   &  0.3  &   11.0 \\
2.050   &   1.0    &   2.0    &    2.0     &     2.0   &     0.3   &      0.79   &   0.8  &   0.8   &    4.4   &   9.3   &  0.3  &   10.9 \\
2.100    &   1.0    &   2.0    &    2.0     &     2.0   &     0.3   &      0.81   &   1.1  &   0.8   &    4.4   &   9.3   &  0.3  &   11.0\\
2.125  &   1.0    &   2.0    &    2.0     &     2.0   &     0.3   &      0.79   &   0.8  &   0.8   &    4.4   &   9.3   &  0.3  &   10.9\\
2.150  &   1.0    &   2.0    &    2.0     &     2.0   &     0.3   &      0.70   &   1.0  &   0.8   &    4.4   &   9.3   &  0.3  &   11.0\\
2.175  &   1.0    &   2.0    &    2.0     &     2.0   &     0.3   &      0.78   &   0.9  &   0.8   &    4.4   &   9.3   &  0.3  &   11.0\\
2.200  &   1.0    &   2.0    &    2.0     &     2.0   &     0.3   &      0.79   &   0.9  &   0.8   &    4.4   &   9.3   &  0.3  &   11.0\\
2.232 &   1.0    &   2.0    &    2.0     &     2.0   &     0.3   &      0.70   &   0.8  &   0.8   &    4.4   &   9.3   &  0.3  &   10.9\\
2.309 &   1.0    &   2.0    &    2.0     &     2.0   &     0.3   &      0.66   &   0.8  &   0.9   &    1.2   &   0.8   &  0.3  &   4.1\\
2.386 &   1.0    &   2.0    &    2.0     &     2.0   &     0.3   &      0.63   &   0.8  &   0.9   &    1.2   &   0.8   &  0.3  &   4.1\\
2.396  &   1.0    &   2.0    &    2.0     &     2.0   &     0.3   &      0.67   &   0.6  &   0.9   &    1.2   &   0.8   &  0.3  &   4.1\\
2.644  &   1.0    &   2.0    &    2.0     &     2.0   &     0.3   &      0.60   &   0.7  &   0.9   &    1.2   &   0.8   &  0.3  &   4.1\\
2.646  &   1.0    &   2.0    &    2.0     &     2.0   &     0.3   &      0.62   &   0.7  &   0.9   &    1.2   &   0.8   &  0.3  &   4.1\\
2.900  &   1.0    &   2.0    &    2.0     &     2.0   &     0.3   &      0.53   &   1.0  &   0.9   &    1.2   &   0.8   &  0.3  &   4.2\\
2.950  &   1.0    &   2.0    &    2.0     &     2.0   &     0.3   &      0.55   &   1.0  &   0.9   &    1.2   &   0.8   &  0.3  &   4.2\\
2.981  &   1.0    &   2.0    &    2.0     &     2.0   &     0.3   &      0.53   &   1.0  &   0.9   &    1.2   &   0.8   &  0.3  &   4.2\\
3.000  &   1.0    &   2.0    &    2.0     &     2.0   &     0.3   &      0.55   &   1.1  &   0.9   &    1.2   &   0.8   &  0.3  &   4.2\\
3.020  &   1.0    &   2.0    &    2.0     &     2.0   &     0.3   &      0.55   &   1.0  &   0.9   &    1.2   &   0.8   &  0.3  &   4.2\\
3.080  &   1.0    &   2.0    &    2.0     &     2.0   &     0.3   &      0.54   &   1.0  &   0.9   &    1.2   &   0.8   &  0.3  &   4.2\\

 \hline
  \hline
\end{tabular}
\label{Sys:ksk}
\end{table}

\section{\label{sec:level7}Fit to the lineshapes}
The measured total Born cross sections for $\epem\to\kpkm\piz$ and the Born cross sections for the intermediate process $\epem\to\phi\piz$ are shown in Fig.~\ref{fig1}, and they are  consistent with the previous results from BaBar and SND.
The cross sections for the processes $\epem \to \ksk$ and $\ktsk$ are shown in Fig.~\ref{figk2k}, where a clear peak between 2.1 GeV and 2.2 GeV is present.

\begin{figure}[htbp]
 \centering
\includegraphics[width=0.45\textwidth]{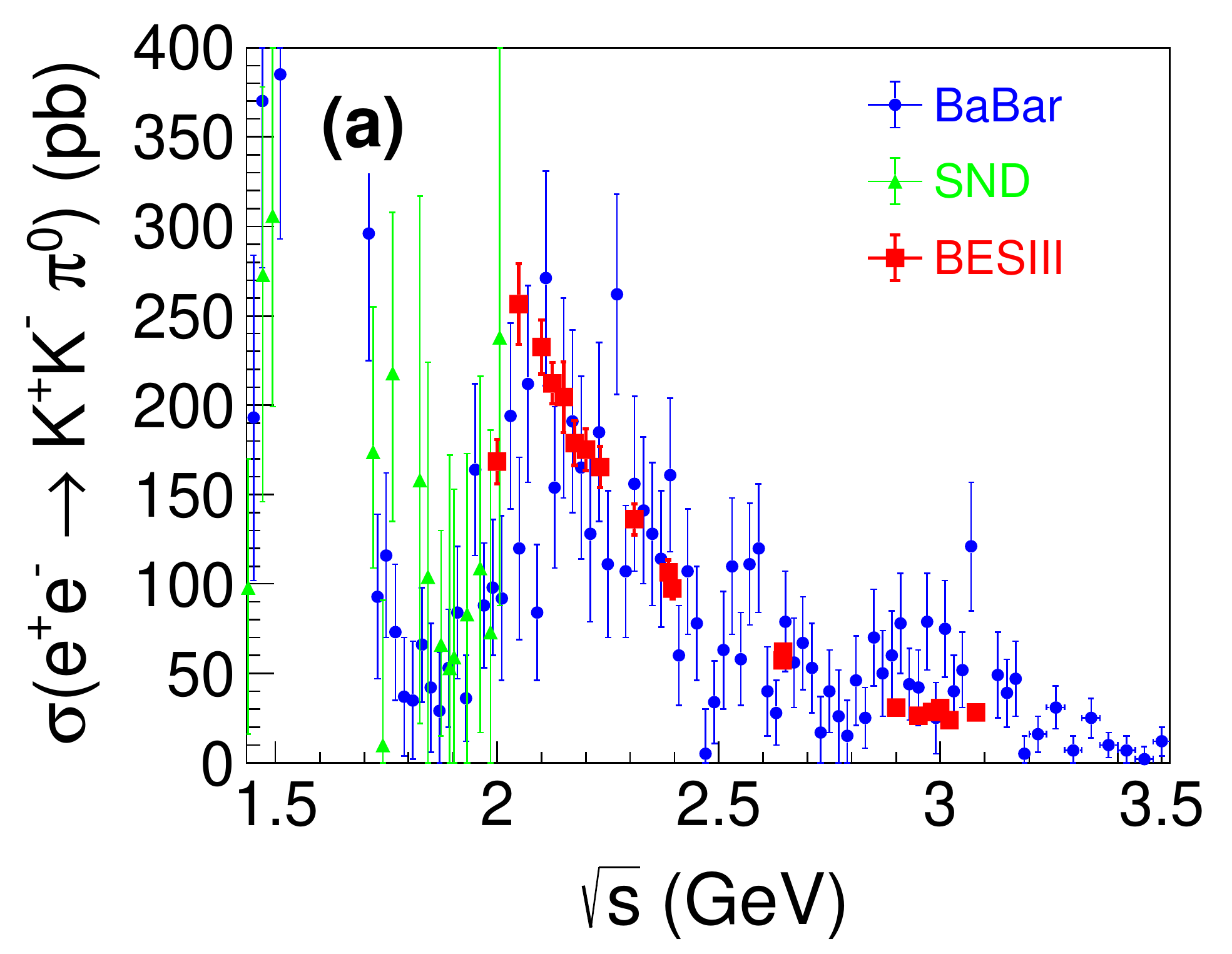}
\includegraphics[width=0.45\textwidth]{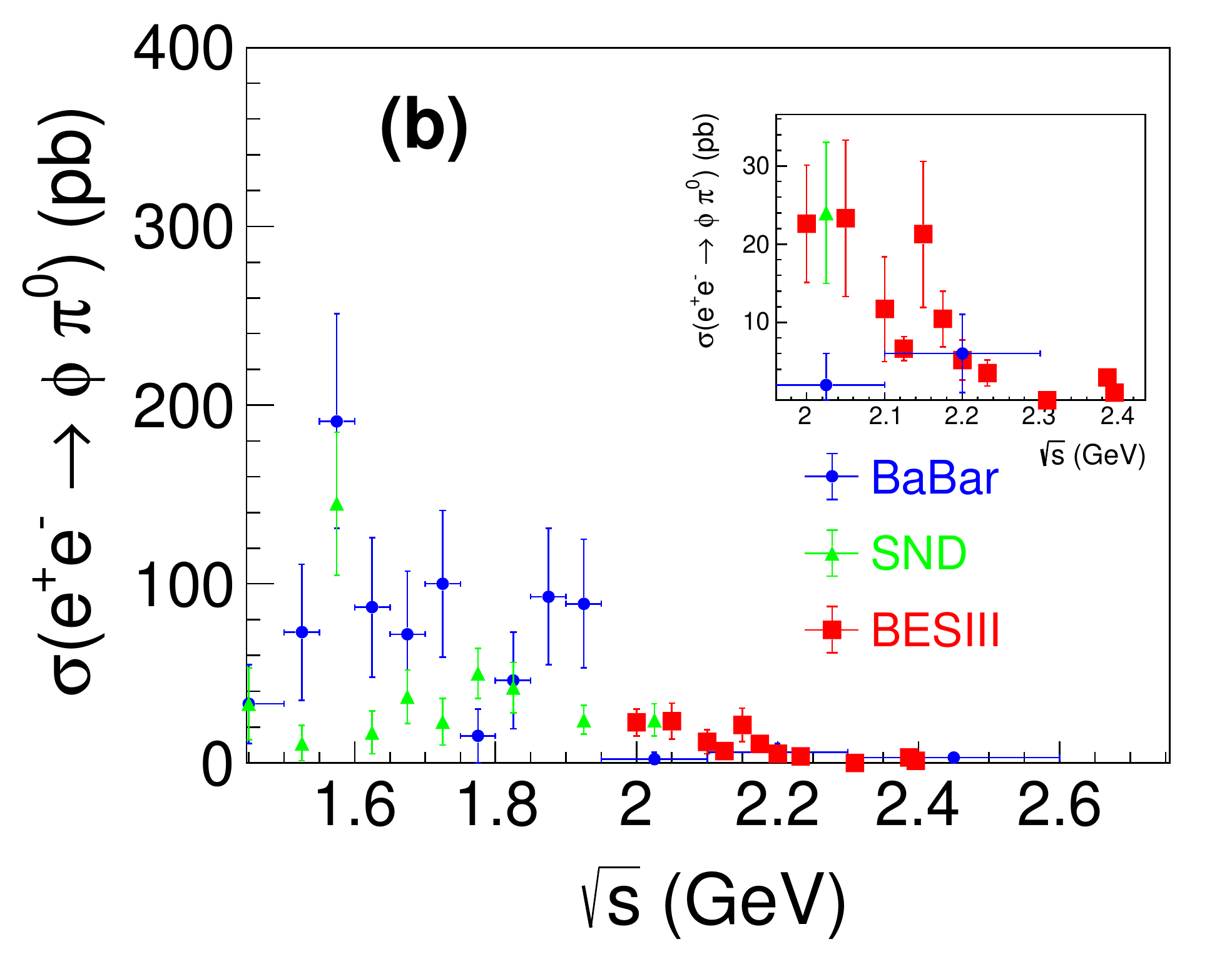}
 \caption{The Born cross sections for (a)~the process $\epem \to \kpkm\piz$ and (b) the intermediate process $\epem \to\phi\piz$.  The red squares are from this analysis; the blue dots and the green triangles are from the BaBar~\cite{babarkkpi} and SND~\cite{sndkkpi} experiments, respectively.}
 \label{fig1}
\end{figure}

 \begin{figure}[htbp]
  \centering
 \includegraphics[width=0.45\textwidth]{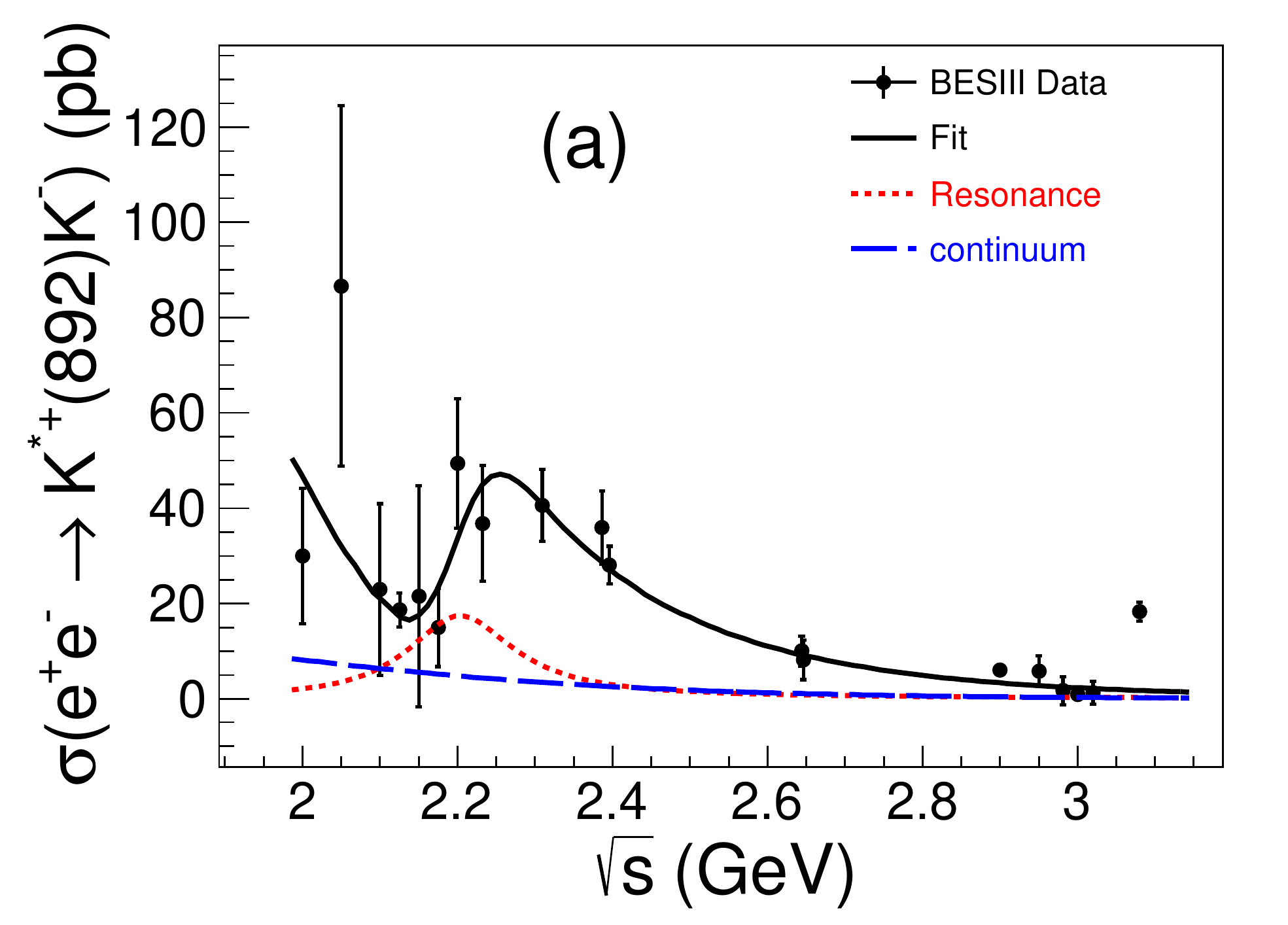}
 \includegraphics[width=0.45\textwidth]{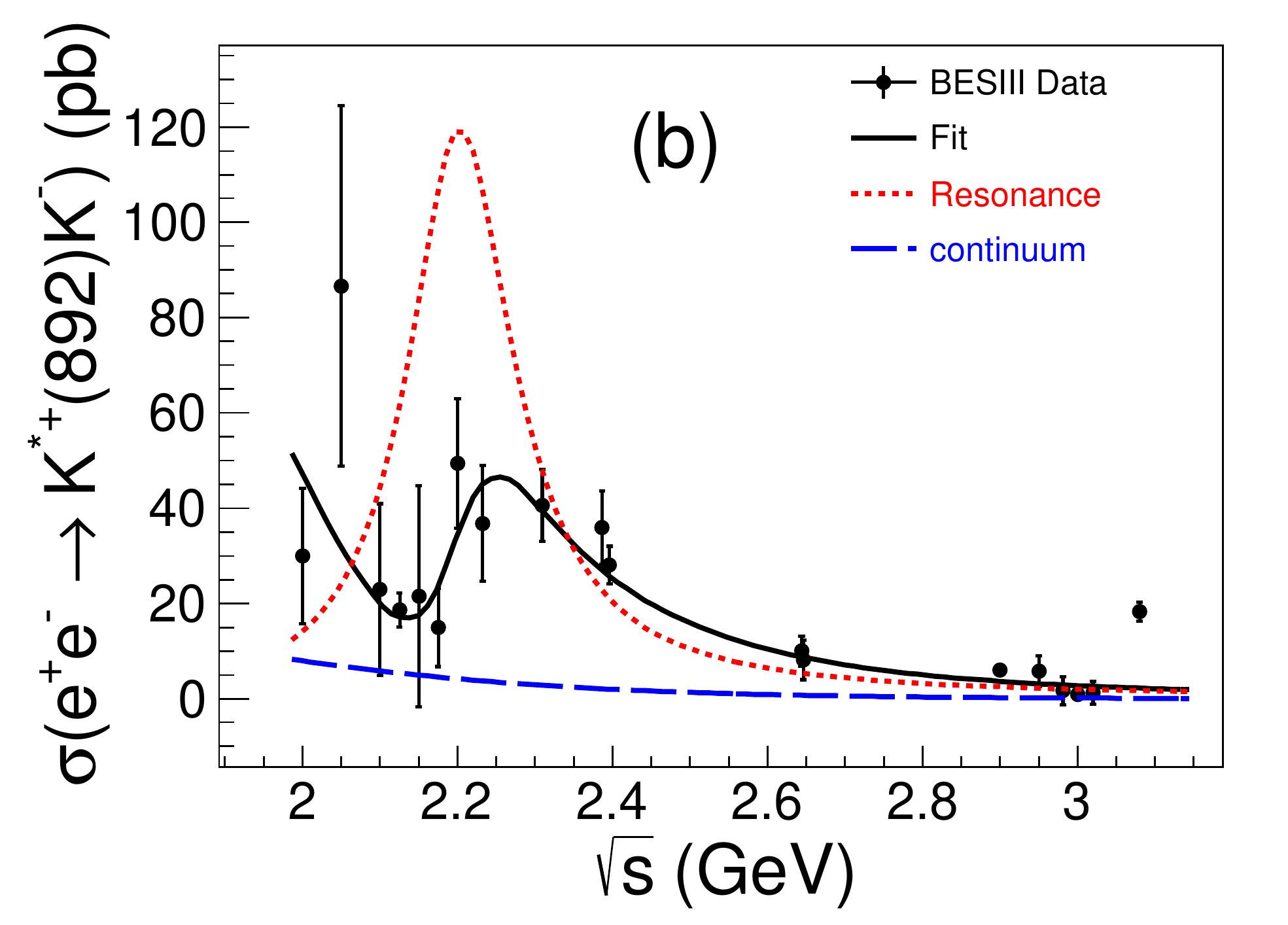}
 \includegraphics[width=0.45\textwidth]{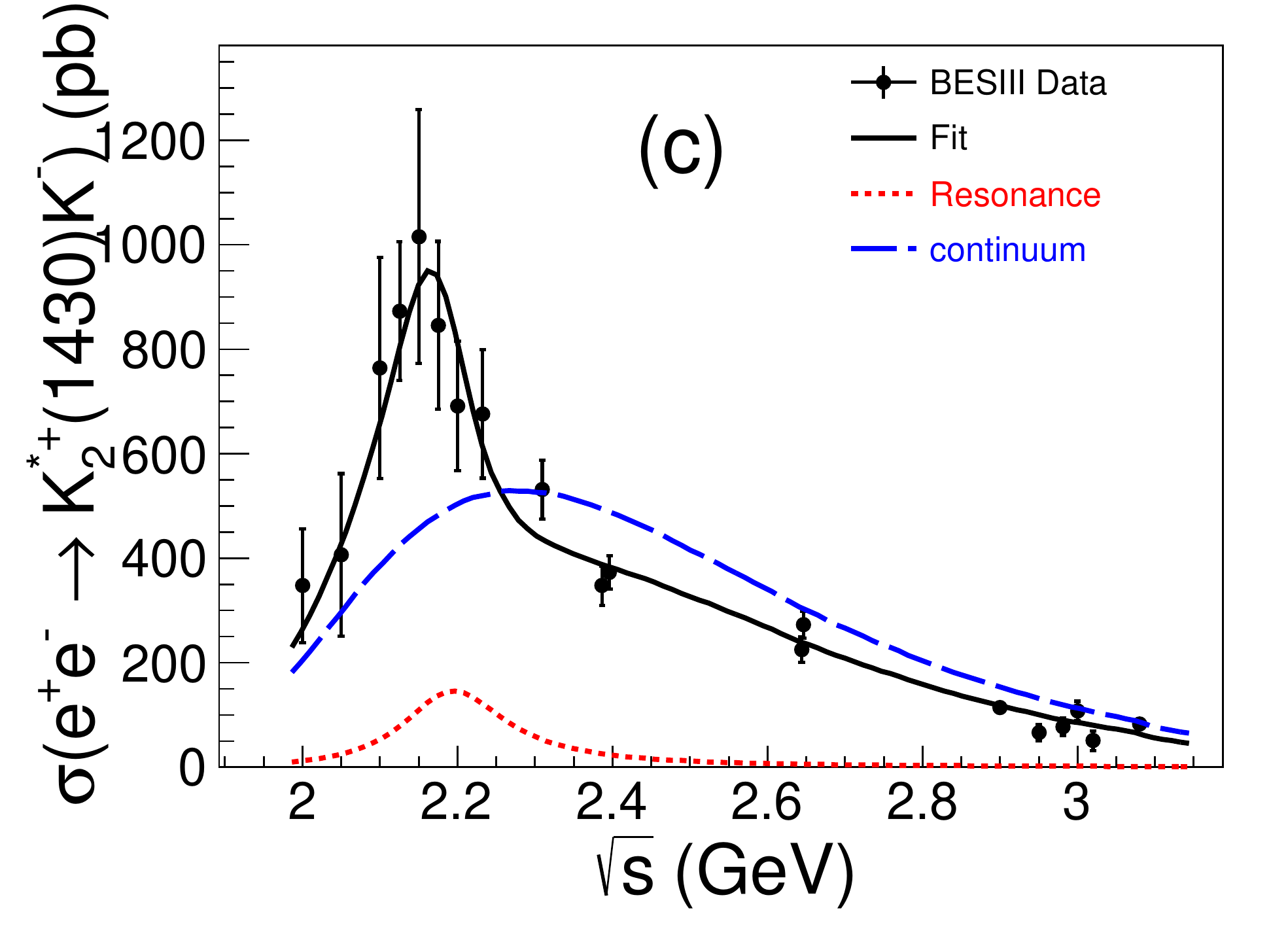}
 \includegraphics[width=0.45\textwidth]{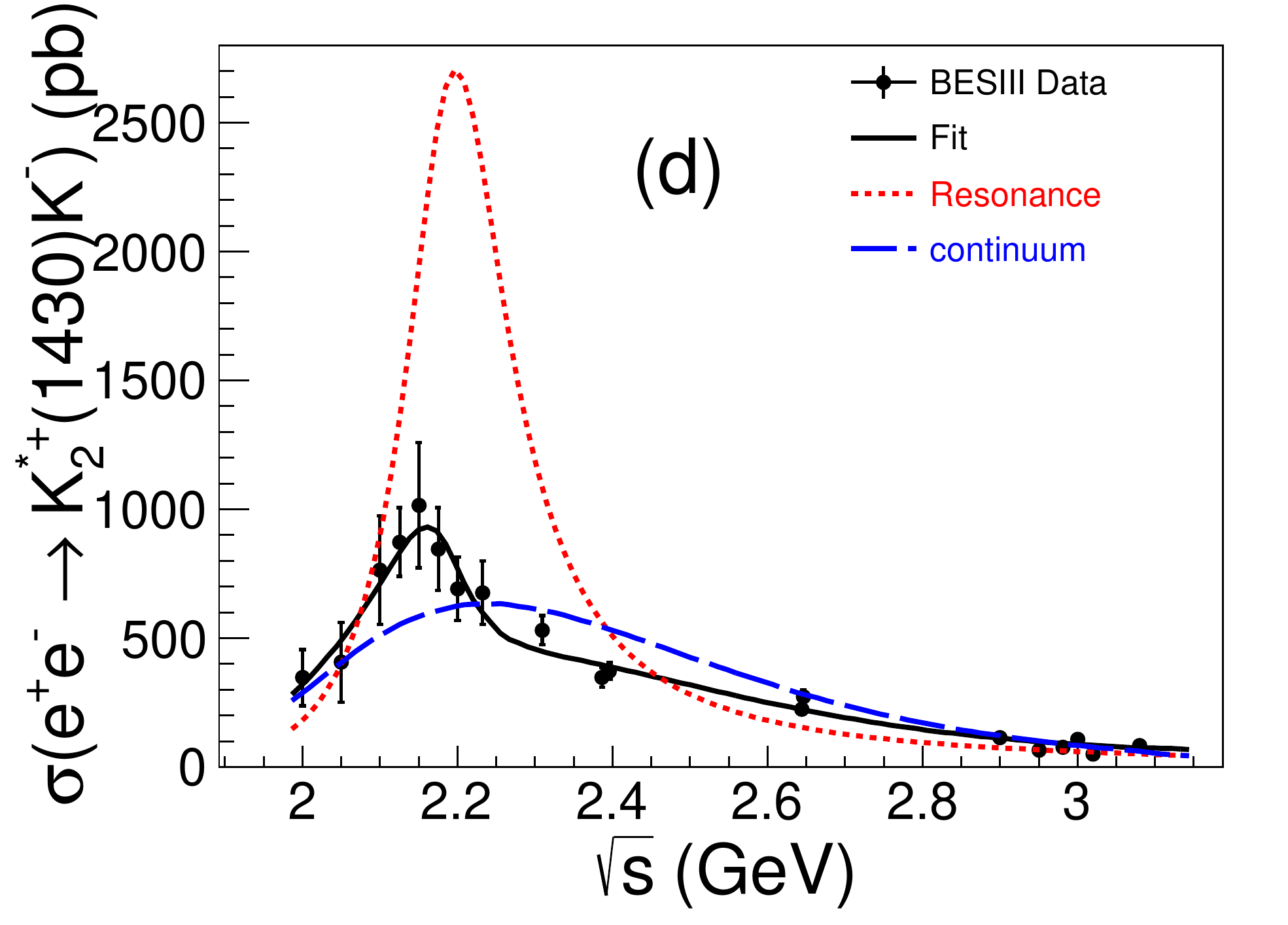}
\caption{ The Born cross section and fit curves for (a) (b)~$\epem \to \ksk$, and for (c) (d)~$\epem \to \ktsk$, corresponding to the two solutions in Table~\ref{09fittable}. Dots with error bars are data, where errors include both statistical and systematic uncertainties. The solid (black) curves represent for the total fit result, the short-dashed (red) curves for the intermediate state and the long-dashed (blue) curves for non-resonant component. }

  \label{figk2k}
 \end{figure}

 To further examine the observed structure in the distributions of the measured cross sections of processes $\epem\to\ktsk$ and $\ksk$, a $\chi^{2}$-fit, incorporating the correlated and uncorrelated uncertainties among different energy points, is performed.
 The fit probability density function (PDF) is a coherent sum of a non-resonant component $f_1$ and a resonant component $f_2$ :
 \begin{equation}
      \mathcal{A} = f_1+ e^{i\varphi} f_2,
 \label{eq1}
 \end{equation}
 where $\varphi$ is the relative phase between the two components.
The non-resonant component includes the contributions from phase space (PHSP) and low-mass resonances, and is described as~\cite{fitnores}.
\begin{equation}
      f_1 = C_0\cdot\sqrt{PS(\sqrt{s})}e^{-p_0(\sqrt{s}-M_{th})},
\label{eq10}
\end{equation}
where $PS(\sqrt{s})$ is the PHSP distribution, $C_0$ and $p_0$ are free parameters, $M_{th}$ is the mass threshold, $M_{th}=m_{K}+m_{K_{2}^{*}(1430)}$ for $\ktsk$ and $M_{th}=m_{K}+m_{K^{*}(892)}$  for $\ksk$. Here, the relative orbital angular momentum in the two-body decay, $L=2$ for the process $\epem\to \ktsk$ and $L=1$ for the process $\epem\to\ksk$, is considered in the $PS(\sqrt{s})$~\cite{PWAtensor} as it follows:

\begin{equation}
     PS(\sqrt{s})  = \int |A_{\ksk,\ktsk}|^{2}d\Phi_{3}, 
\label{eq11}
\end{equation} 
where $A$ is partial wave amplitude in the covariant Rarita-Schwinger tensor formalism~\cite{PWAtensor}, $\Phi_{3}$ is three-body phase space.
 The amplitudes for $\ktsk$ and $K^{*+}(892)K^{-}$ are described as:
\begin{equation}
     A_{\mu,\ktsk}=-\epsilon_{\mu\nu\lambda\sigma}p^{\sigma}_{\psi}\widetilde{T}^{(2)\nu\alpha}_{\ktsk}\cdot f^{K^{*+}_{2}}_{K^{+}\piz} \cdot \widetilde{t}^{(2)\lambda}_{K^{+}\piz \alpha},
\label{eq12}
\end{equation}

\begin{equation}
     A_{\mu,\ksk}=-\epsilon_{\mu\nu\lambda\sigma}p^{\sigma}_{\psi}\widetilde{T}^{(1)\nu}_{\ksk}\cdot f^{K^{*+}}_{K^{+}\piz} \cdot \widetilde{t}^{(1)\lambda}_{K^{+}\piz},
\label{eq12}
\end{equation}
where $\widetilde{T}, \widetilde{t}$ are the covariant tensors, $f$ is a Breit-Wigner propagator~\cite{PWAtensor}, $\epsilon_{\mu\nu\lambda\sigma}$ is the Levi-Civita symbol, the other operators are found in reference~\cite{PWAtensor}.

The resonant amplitude $f_2$ is described with a BW function,
\begin{equation}
      f_2 = \frac{\sqrt{12\pi \mathcal{B}r \Gamma^{e^{+}e^{-}}_R \Gamma_R}}{s-M_R^{2}+iM_R\Gamma(\sqrt{s})}\cdot\sqrt{\frac{PS(\sqrt{s})}{PS(M_R)}},
\label{eq2}
\end{equation}

\begin{equation}
\Gamma(\sqrt{s})=\Gamma_R\frac{PS(\sqrt{s})}{PS(M_R)},
\label{eq21}
\end{equation}
where $M_R$ is the mass of the resonance,
$\Gamma_R$ is the constant width,
$\Gamma^{\epem}_R $ is its partial width to $\epem$,
and $\mathcal{B}r$ is the decay branching fraction to a given final state.

 A simultaneous fit, assuming the same resonant structures in the $\epem \to \ktsk$ and $\ksk$ processes,
 is performed to the measured cross sections.
 In the fit, $M_R$ and $\it \Gamma_R$ are shared parameters between the two processes and floating, while the production $\mathcal{B}r \Gamma^{\epem}_R$ and the relative phase angle $\varphi$ are independent between two processes.
 The fit yields two solutions with equal fit quality and identical $M_R=(2208 \pm 19)~\mevcc$ and $\Gamma_R=(168 \pm 24)~\mev$.
 The fit curves are shown in Fig.~\ref{figk2k}, and the results are  summarized in Table~\ref{09fittable}.
The overall significance of this resonance is estimated to be 7.6$\sigma$ for the $\epem \to \ktsk$ and $\ksk$ processes, by comparing the change of $\chi^{2}$ ($\Delta\chi^{2}$), with and without the resonant structure in the fit and taking the change of degrees of freedom into account. 
The significances of the resonant state for the two individual processes are also estimated and summarized in Table~\ref{09fittable}.

\begin{table}[htbp]
\begin{center}
\scriptsize
\caption{A summary of fit results.}
\begin{tabular}{l |c | c |c |c}
\hline\hline
channel       &   & $\mathcal{B}_{r} \Gamma^{e^{+}e^{-}}_{R}$ (eV) &  $\varphi$ (rad)  &  Sig. ($\sigma$) \\
\hline
\multirow{2}{*}{$\ktsk$}  &  solution1 &  8.2   $\pm$ 2.2   &   2.9 $\pm$ 0.3  & \multirow{2}{*}{5.7}\\
                          &  solution2 &  142.4 $\pm$ 18.7  &   4.8 $\pm$ 0.1  &   \\
\hline
\multirow{2}{*}{$\ksk$}     &  solution1 &  1.1 $\pm$ 0.3 & 3.0 $\pm$ 0.3  & \multirow{2}{*}{4.8}\\
                          &  solution2 &  6.5 $\pm$ 0.9 & 1.8 $\pm$ 0.1  &   \\
 \hline
  \hline
\end{tabular}
\label{09fittable}
\end{center}
\end{table}

  The systematic uncertainties on the resonant  parameters come from the absolute c.m.\ energy measurement, the measured cross section, and the fit procedure.
  The uncertainty of the c.m.~energy from BEPCII is small and is ignored in the determination of the parameters of the structure.
  The statistical and systematic uncertainties of the measured cross section are incorporated in the fit, thus no further uncertainty is necessary.
  The uncertainties associated with the fit procedure include those from the signal model.
  To assess the systematic uncertainty associated with the signal model, an alternative BW function  with constant width is implemented in the fit, and the resulting differences of 16.3 $\mevcc$  and 37.5~$\mev$ in mass and width, respectively, are considered as the related systematic uncertainties.
The uncertainty of the parametrization of the continuum contribution is estimated by changing the term $e^{-p_0(\sqrt{s}-M_{th})}$ in Eq.~(\ref{eq10})  with $1/s^{n}$, where $n$ is a free parameter. The differences of the obtained mass and  width, which are 17.2 $\mevcc$ and 10.0 MeV, respectively, are assigned as the corresponding systematic uncertainties. 
The overall systematic uncertainties are the quadratic sum of the individual ones, 23.6~$\mevcc$ and 38.8~$\mev$ for the mass and width, respectively.

\section{Summary}

In summary,  a PWA of the process $\epem \to \kpkm\piz$ is performed for nineteen data samples with c.m.~energies between 2.000 and 3.080~$\gev$ and  a total integrated luminosity of 648~$\ipb$ taken by the BESIII detector.
The Born cross section of $\epem\to \kpkm\piz$, as well as those for the intermediate processes  $\epem\to\phi \piz$, $K^{*+}(892)K^{-}$ and $\ktsk$, are measured by performing a PWA on each data sample individually with two baseline solutions according to its c.m.~energies.
The cross section for $\epem \to \kpkm\piz$ and $\phi \piz$ is measured with improved precision and is consistent with those measured by the BaBar experiment.
A structure is observed in the cross section of the intermediate processes $\epem\to \ksk$ and $\ktsk$, and
by performing a simultaneous $\chi^2$ fit, the two solutions which were obtained confirmed
a resonance with mass $M_R$ = (2208 $\pm$ 19 $\pm$ 24)~$\mevcc$, width $\Gamma_R$ = (168 $\pm$ 24 $\pm$ 39)~$\mev$, and a significance of 7.6$\sigma$, where the uncertainties are statistical and systematic, respectively.
The observed resonance is directly produced in $\epem$ collisions, thus a $J^{PC}=1^{--}$ is assigned.
Comparing to the vector mesons listed in the PDG~\cite{pdg}, the mass of the observed resonance is close to those of $\phi(2170)$, $\rho(2150)$ and $\omega(2290)$, and its width 
is consistent with that of $\phi(2170)$ within uncertainties, but deviates from those of $\rho(2150)$ and $\omega(2290)$ by more than 3$\sigma$.

Assuming the observed structure is $\phi(2170)$, 
the relative branching ratio 
$$R= \frac{\mathcal{B}r(\phi(2170) \to \ktsk)}{\mathcal{B}r(\phi(2170) \to \ksk)}$$ is calculated to be $7.5 \pm 2.9$ and $21.9 \pm 4.2$ for solution 1 and solution 2, respectively.
It is noticeable that the branching fraction of $\phi(2170) \to \ktsk$ is significantly larger than that of $\phi(2170) \to \ksk$, which could provide more insight into the nature of $\phi(2170)$.


\section{Acknowledgments}
The BESIII collaboration thanks the staff of BEPCII and the IHEP computing center and the supercomputing center of USTC for their strong support. This work is supported in part by National Key R\&D Program of China under Contracts Nos. 2020YFA0406400, 2020YFA0406300; National Natural Science Foundation of China (NSFC) under Contracts Nos. 11335008, 11625523, 11635010, 11735014, 11822506, 11835012, 11935015, 11935016, 11935018, 11961141012, 12022510, 12025502, 12035009, 12035013, 12061131003, 11705192, 11975118,11950410506, 12061131003; the Natural Science Foundation of Hunan Province under Contract Nos. 2019JJ30019 and 2020RC3054; the Chinese Academy of Sciences (CAS) Large-Scale Scientific Facility Program; Joint Large-Scale Scientific Facility Funds of the NSFC and CAS under Contracts Nos. U1732263, U1832207, U1832103, U2032111, U2032105; CAS Key Research Program of Frontier Sciences under Contract No. QYZDJ-SSW-SLH040; 100 Talents Program of CAS; INPAC and Shanghai Key Laboratory for Particle Physics and Cosmology; ERC under Contract No. 758462; European Union Horizon 2020 research and innovation programme under Contract No. Marie Sklodowska-Curie grant agreement No 894790; German Research Foundation DFG under Contracts Nos. 443159800, Collaborative Research Center CRC 1044, FOR 2359, FOR 2359, GRK 214; Istituto Nazionale di Fisica Nucleare, Italy; Ministry of Development of Turkey under Contract No. DPT2006K-120470; National Science and Technology fund; Olle Engkvist Foundation under Contract No. 200-0605; STFC (United Kingdom); The Knut and Alice Wallenberg Foundation (Sweden) under Contract No. 2016.0157; The Royal Society, UK under Contracts Nos. DH140054, DH160214; The Swedish Research Council; U. S. Department of Energy under Contracts Nos. DE-FG02-05ER41374, DE-SC-0012069.

\vspace{0.5cm}
{\bf BESIII collaboration}\\

\input{author}
\end{document}

%% file: author.tex
\noindent
M.~Ablikim$^{1}$, M.~N.~Achasov$^{10,b}$, P.~Adlarson$^{68}$, S. ~Ahmed$^{14}$, M.~Albrecht$^{4}$, R.~Aliberti$^{28}$, A.~Amoroso$^{67A,67C}$, M.~R.~An$^{32}$, Q.~An$^{64,50}$, X.~H.~Bai$^{58}$
% ...

\vspace{0.2cm} {\it
\noindent
$^{1}$ Institute of High Energy Physics, Beijing 100049, People's Republic of China\\
% ...
$^{4}$ Bochum Ruhr-University, D-44780 Bochum, Germany\\
% ...
$^{10}$ G.I. Budker Institute of Nuclear Physics SB RAS (BINP), Novosibirsk 630090, Russia\\
% ...
$^{14}$ Helmholtz Institute Mainz, Staudinger Weg 18, D-55099 Mainz, Germany\\
% ...
$^{28}$ Johannes Gutenberg University of Mainz, Johann-Joachim-Becher-Weg 45, D-55099 Mainz, Germany\\
% ...
$^{32}$ Liaoning Normal University, Dalian 116029, People's Republic of China\\
% ...
$^{50}$ State Key Laboratory of Particle Detection and Electronics, Beijing 100049, Hefei 230026, People's Republic of China\\
% ...
$^{58}$ University of Jinan, Jinan 250022, People's Republic of China\\
% ...
$^{64}$ University of Science and Technology of China, Hefei 230026, People's Republic of China\\
% ...
$^{67}$ University of Turin and INFN, (A)University of Turin, I-10125, Turin, Italy; (B)University of Eastern Piedmont, I-15121, Alessandria, Italy; (C)INFN, I-10125, Turin, Italy\\
$^{68}$ Uppsala University, Box 516, SE-75120 Uppsala, Sweden\\
% ...
$^{b}$ Also at the Novosibirsk State University, Novosibirsk, 630090, Russia\\
% ...
}

%% file: kkpi0.bbl
\begin{thebibliography}{99}
 
   \bibitem{pdg} P.~A.~Zyla {\it et al.} (Particle Data Group), Prog.\ Theor.\ Exp.\ Phys. {\bf 2020}, 083C01 (2020).
   \bibitem{Y2170Babar} B.~Aubert {\it et al.} (BaBar Collaboration),  Phys.\ Rev.\ D {\bf 74}, 091103(R) (2006).
   \bibitem{Y2170Babar1} B.~Aubert {\it et al.} (BaBar Collaboration), Phys.\ Rev.\ D {\bf 76}, 012008 (2007).
   \bibitem{Y2170belle} C.~P.~Shen {\it et al.} (Belle Collaboration), Phys.\ Rev.\ D  {\bf 80}, 031101(R) (2009).
   \bibitem{Y2170Bes} M.~Ablikim {\it et al.} (BES Collaboration), Phys.\ Rev.\ Lett. {\bf 100}, 102003 (2008).
   \bibitem{Y2170Bes3} M.~Ablikim {\it et al.} (BESIII Collaboration), Phys.\ Rev.\ D {\bf 91}, 052017 (2015).
   \bibitem{2017bes32} M.~Ablikim {\it et al.} (BESIII Collaboration), Phys.\ Rev.\ D {\bf 99}, 012014 (2019).
   \bibitem{bes3kk} M. Ablikim {\it et al}. (BESIII Collaboration),\ Phys.\ Rev.\ D {\bf 99}, 032001 (2019).
   \bibitem{bes3phikk}  M. Ablikim {\it et al}. (BESIII Collaboration),\ Phys.\ Rev. D{\bf 100}, 032009 (2019).
   \bibitem{bes3kkpi0pi0} M. Ablikim et al. (BESIII Collaboration),\ Phys.\ Rev.\ Lett.{\bf 124}, 112001 (2020).
   \bibitem{bes3omegaeta} M. Ablikim et al. (BESIII Collaboration),\ Phys.\ Lett.\ B {\bf 813}, 136059 (2021).
   \bibitem{bes3phietap} M. Ablikim {\it et al}. (BESIII Collaboration),\ Phys.\ Rev.\ D {\bf 102}, 012008 (2020).
   \bibitem{bes3phieta} M. Ablikim {\it et al}. (BESIII Collaboration),\ Phys.\ Rev.\ D {\bf 104}, 032007 (2021).

   \bibitem{strange} T.~Barnes, {\it et al.},\ Phys.\ Rev.\ D {\bf 68} 054014 (2003).
   \bibitem{2017ding} G.~J.~Ding and M.~L.~Yan,\ Phys.\ Lett.\ B {\bf 657}, 49 (2007).
   \bibitem{2017wang} X.~Wang {\it et al.},\ Phys.\ Rev.\ D {\bf 85}, 074024 (2012).
   \bibitem{2017afonin} S.~S.~Afonin and I.~V.~Pusenkov,\ Phys.\ Rev.\ D {\bf 90}, 094020 (2014).
   \bibitem{2017ding2} G.~J.~Ding and M.~L.~Yan,\ Phys.\ Lett.\ B {\bf 650}, 390 (2007).
   \bibitem{hybrid2} P.~R.~Page, E.~S.~Swanson, and A.~P.~Szczepaniak,\ Phys.\ Rev.\ D {\bf 59}, 034016 (1999).
   \bibitem{2017wang2} Z.~G.~Wang,\ Nucl.\ Phys.\ A {\bf 791}, 106 (2007).
   \bibitem{2017chen} H.~X.~Chen {\it et al.},\ Phys.\ Rev.\ D {\bf 78}, 034012 (2008).
   \bibitem{2019ke} H.~W.~Ke and X.~Q.~Li,\ Phys.\ Rev.\ D {\bf 99}, 036014 (2019).
   \bibitem{2017drenska} N.~V.~Drenska, R.~Faccini and A.~D.~Polosa,\ Phys.\ Lett.\ B {\bf 669}, 160 (2008).
   \bibitem{2017zhao} L.~Zhao {\it et al.},\ Phys.\ Rev.\ D {\bf 87}, 054034 (2013).
   \bibitem{2017deng} C.~Deng {\it et al.},\ Phys.\ Rev.\ D {\bf 88}, 074007 (2013).
   \bibitem{2017dong} Y.~Dong {\it et al.},\ Phys.\ Rev.\ D {\bf 96}, 074027 (2017).
   \bibitem{2017oset} A.~Martinez Torres {\it et al.}, Phys. Rev. D {\bf 78}, 074031 (2008);
   S.~Gomez-Avila, M.~Napsuciale and E.~Oset,\ Phys.\ Rev.\ D {\bf 79}, 034018 (2009).
   \bibitem{babarkkpi} B. Aubert {\it et al}. (BABAR Collaboration),\ Phys.\ Rev.\ D {\bf 77}, 092002 (2008).

\bibitem{hybridchen} Y.~Ma {\it et al.},\ Chin.\ Phys.\ C {\bf 45}, 013112 (2021). 

\bibitem{bes3} M.~Ablikim {\it et al.} (BESIII Collaboration), Nucl.\ Instrum.\ Meth.\ A {\bf 614}, 3 (2010).
\bibitem{bepc2} C.~H.~Yu {\it et al.}, Proceedings of IPAC2016, Busan, Korea, 2016, doi:10.18429/JACoW-IPAC2016-TUYA01.
\bibitem{Ablikim:2019hff} M.~Ablikim {\it et al.} (BESIII Collaboration), Chin.\ Phys.\ C {\bf 44}, 04000(2020).
\bibitem{geant4} S.~Agostinelli {\it et al.} ({\sc Geant4} Collaboration), Nucl.\ Instrum.\ Meth. A {\bf 506}, 250 (2003).
\bibitem{conexc} R.~G.~Ping,\ Chin.\ Phys.\ C {\bf 38}, 083001 (2014).
\bibitem{babayaga} G.~Balossini {\it et al.},\ Nucl.\ Phys.\ B {\bf 758}, 227 (2006).
\bibitem{lumarlw} B.~Andersson and H.~Hu, arXiv:hep-ph/9910285.
\bibitem{bestwogam} M.~Ablikim, {\it et al.} (BESIII Collaboration), Chin.\ Phys.\ C {\bf 41}, 063001 (2017).
\bibitem{PWAframe} N.~Berger, B.~J.~Liu, and J.~K.Wang,\ J.\ Phys.\ Conf.\ Ser. 219, 042031 (2010).
\bibitem{PWAtensor} B.~S.~Zou and D.~V.~Bugg,\ Eur.\ Phys.\ J.\ A {\bf 16}, 537 (2003).
\bibitem{PWA:sigma1} M.~Ablikim, {\it et al.} (BESII Collaboration),\ Phys.\ Lett. B {\bf 598}, 149 (2004).
\bibitem{PWA:sigma2} M. Ablikim, {\it et al.} (BESII Collaboration),\ Phys.\ Lett. B {\bf 645}, 19 (2007).
\bibitem{PWA:minuit1} F.~James and M.\ Roos,\ Comput.\ Phys.\ Commun. 10, 343 (1975).
\bibitem{phipiextic} F.~E.~Close and H.~J.~Lipkin,\ Phys.\ Lett.\ B {\bf 196}, 245 (1987).
\bibitem{VR} E.~A.~Kuraev and V.~S.~Fadin,\ Sov.\ J.\ Nucl.\ Phys. {\bf 41}, 466 (1985).
\bibitem{VP} S.~Actis {\it et al}., Eur.\ Phys.\ J.\ C {\bf 66}, 585 (2010).
\bibitem{lum} M.~Ablikim {\it et al.} (BESIII Collaboration), Chin.\ Phys.\ C {\bf 41}, 063001 (2017).
\bibitem{trackerror} M.~Ablikim {\it et al.} (BESIII Collaboration), Phys.\ Rev.\ D {\bf 99}, 032001 (2019).
\bibitem{photonerror} M.~Ablikim {\it et al.} (BESIII Collaboration), Phys.\ Rev.\ D {\bf 81}, 052005 (2010).
\bibitem{helixsys} M.~Ablikim {\it et al.} (BESIII Collaboration), Phys.\ Rev.\ D {\bf 87}, 012002 (2013).
\bibitem{fitnores} L.~Peter,\ Phys.\ Rev.\ D {\bf 98}, 113011 (2018).
\bibitem{sndkkpi} M. N. Achasov {\it et al.} (SND Collaboration), \ Eur.\ Phys.\ J.\ C {\bf 80}. 1139(2020).

\end{thebibliography}
